\documentclass[prd,preprintnumbers,superscriptaddress,nofootinbib,twocolumn]{revtex4-2}
\usepackage{epsfig}
\usepackage{amsmath}
\usepackage{amsfonts}
\usepackage{float}
\usepackage{amssymb}
\usepackage{slashed}
\usepackage{color}
\usepackage{pbox}
\usepackage{subfigure}
\usepackage[colorlinks,citecolor=blue, linkcolor = blue, urlcolor = blue]{hyperref}
\usepackage{tabularx}
\usepackage{titlesec}
\usepackage{scrextend}
\usepackage[normalem]{ulem}
\usepackage{orcidlink}
\usepackage{comment}
\usepackage{MnSymbol}

\def\lsim{\raise0.3ex\hbox{$\;<$\kern-0.75em\raise-1.1ex\hbox{$\sim\;$}}}
\def\gsim{\raise0.3ex\hbox{$\;>$\kern-0.75em\raise-1.1ex\hbox{$\sim\;$}}}

\def\lsim{\lesssim}

\def\trh{T_{\rm RH}}
\def\gsim{\gtrsim}
\def\bal#1\eal{\begin{align}#1\end{align}}

\graphicspath{{Figures/}}

\begin{document}


\title{Viability of post-inflationary freeze-in with precision cosmology}

\author{Anirban Biswas}
\email{anirban.biswas.sinp@gmail.com}
\affiliation{Department of Physics, Gaya College (A constituent
unit of Magadh University, Bodh Gaya), Gaya 823001, India}
\author{Sougata Ganguly
\orcidlink{0000-0002-8742-0870}}
\email{sganguly0205@ibs.re.kr}
\affiliation{Particle Theory  and Cosmology Group (PTC),
Center for Theoretical Physics of the Universe (CTPU), \\
Institute for Basic Science, Daejeon 34126, Republic of Korea}
\author{Dibyendu Nanda\orcidlink{0000-0002-7768-7029}}
\email{dnanda@het.phys.sci.osaka-u.ac.jp}
\affiliation{Department of Physics, Osaka University, Toyonaka, Osaka 560-0043, Japan}
\author{Sujit Kumar Sahoo\orcidlink{0000-0002-9014-933X}}
\email{ph21resch11008@iith.ac.in}
\affiliation{Department of Physics, Indian Institute of Technology Hyderabad, Kandi, Telangana-502285, India.}
\begin{abstract}
Prediction of inflationary observables from the temperature fluctuation of Cosmic Microwave Background (CMB) can play a pivotal role in predicting the reheating dynamics in the early universe. In this work, we highlight how the inflationary observables, in particular the spectral index $n_s$, can play a potential role in constraining the post-inflationary dark matter (DM) production. We demonstrate a novel way of constraining the non-thermal production of DM via UV freeze-in which is otherwise elusive in terrestrial experiments. We consider a scenario in which DM is produced from this thermal plasma via a dimension-five operator. 
The mutual connection between $n_s$ and relic density of DM via the reheating temperature, $T_{\rm RH}$, enables us to put constraints on the DM parameter space. For the minimal choice of the inflationary model parameters and DM mass between $1\,\rm MeV$ to $1\,\rm TeV$, we found that Planck alone can exclude the cut-off scale of the dimension-five operator $\Lambda \lesssim 10^{12}\,\rm GeV$  which is significantly stronger than any other existing constraints on such minimal scenario. If we impose the combined prediction form Planck and recently released data by ACT, the exclusion limit can reach up to the Planck scale for TeV-scale dark matter. 
\end{abstract}
\preprint{CTPU-PTC-25-19}
\preprint{OU-HET 1277}
\maketitle

\section{Introduction} 
\label{sec:intro}
The origin of dark matter (DM) remains one of the most profound mysteries of modern particle physics and cosmology, with numerous theoretical frameworks attempting to explain its existence. The conventional Weakly Interacting Massive Particle (WIMP) paradigm, extensively studied in the literature, posits that DM particles were once in thermal equilibrium with the Standard Model (SM) bath before decoupling or freezing out \cite{Arcadi:2017kky, Roszkowski:2017nbc,Arcadi:2024ukq}. WIMPs have attracted considerable interest, offering several avenues for experimental verification, including direct detection, indirect detection, and collider searches. Despite these extensive efforts for more than a decade, no clear evidence for WIMPs has emerged, and the resulting experimental constraints have considerably limited the available parameter space \cite{SuperCDMS:2015eex, XENON:2018voc,LZ:2022lsv, Lattaud:2022jnq, XENON:2023cxc, MAGIC:2016xys, HESS:2022ygk}.

As an alternative to WIMPs, the Feebly Interacting Massive Particle (FIMP) has recently attracted growing attention. In this framework, DM particles are produced through extremely weak interactions with the bath particles and never achieve equilibrium \cite{Kusenko:2006rh, McDonald:2008ua, Hall:2009bx, Elahi:2014fsa,Bernal:2017kxu}.  As a result of the feeble interaction with the visible sector, FIMPs can evade the direct detection bounds while satisfying the cosmological abundance of DM. The requirement of an extremely suppressed interaction strength between FIMP and the SM sector can be realized either by introducing tiny dimensionless couplings ($\sim 10^{-10}$) or through non-renormalizable operators suppressed by a high cut-off scale. Models where the dark sector communicates with the SM via renormalizable interactions with tiny couplings are referred to as infrared (IR) freeze-in \cite{Hall:2009bx}. In contrast, in the ultraviolet (UV) freeze-in scenario, DM production is dominated by non-renormalizable operators \cite{Elahi:2014fsa}. In this case, the DM yield is proportional to the reheating temperature ($\trh$), which marks the onset of the radiation-dominated era after inflation \cite{Giudice:2000ex, Kofman:1997yn}. Thus post-inflationary reheating phase plays a crucial role in setting the DM relic produced via UV freeze-in. In this context, it is important to understand that the allowed range of reheating temperature can span many orders of magnitude. While the lower bound on $\trh$ is set by successful big bang nucleosynthesis (BBN), requiring it to be $\geq 4 $ MeV \cite{Ichikawa:2005vw, Kawasaki:2000en, Barbieri:2025moq}, the upper bound can be derived from the maximum allowed  energy density of the universe after the inflation era $\rho_{\rm max}\sim 10^{66} \text{ GeV}^4$ \cite{Baumann:2009ds}, as suggested by Planck data \cite{Planck:2018vyg}.

For a given inflationary model, the predictions of key observables such as the spectral index ($n_s$) and  tensor-to-scalar ratio ($r$) crucially depend on the details of reheating, especially on the number of e-folds from the end of the inflation to the beginning of the radiation domination era \cite{Cook:2015vqa, Ueno:2016dim, Drewes:2017fmn, Maity:2018dgy, Maity:2018exj, Drewes:2019rxn, Haque:2020zco, DiMarco:2021xzk}. From the current cosmological data, both $n_s$ and $r$ are now tightly constrained. The Planck 2018 data \cite{Planck:2018jri} reported $n_s = 0.965 \pm 0.004$ at 95 \% confidence level (CL). More recently, the Atacama Cosmology Telescope (ACT) collaboration has released their results \cite{ACT:2025tim, ACT:2025fju}, finding $n_s = 0.9666 \pm 0.0077$ from ACT data alone, consistent with the Planck result. Interestingly, when combining both ACT and Planck datasets, due to opposite correlations between $n_s$ and $\Omega_b  h^2$ in the two experiments, the fit shifts to a higher value, yielding $n_s = 0.9709 \pm 0.0038$. This has led to several discussions on possible implications and interpretations in the context of inflationary models \cite{Kallosh:2025rni, Pan:2025psn, Dioguardi:2025vci, Brahma:2025dio, Gialamas:2025kef, Antoniadis:2025pfa, Gao:2025onc, Wang:2025zri, Yin:2025rrs, Liu:2025qca, Gialamas:2025ofz, McDonald:2025odl}. On the contrary, the
prediction of the upper limit of $r$ at 95 \%
CL from Planck and Planck + ACT are
0.036 and 0.038 respectively.

Given that UV freeze-in DM production crucially depends on the reheating temperature, and that $\trh$ is in turn constrained by precise prediction of $n_s$ and $r$, it becomes imperative to explore the connection between DM production and $n_s$ in the light of current cosmological data. Here we investigate the interplay between the DM relic abundance and inflationary observables within the framework of the $\alpha$-attractor model of inflation \cite{Kallosh:2013lkr, Kallosh:2013hoa, Kallosh:2013yoa, Kallosh:2013pby, Kallosh:2013maa, Kallosh:2013yoa, Galante:2014ifa}. In our framework, SM particles are produced from the coherent oscillation of the inflaton and DM  is subsequently produced via UV freeze-in through a dimension-five operator.
While previous studies have explored the connection between the $\trh$ and production of non-thermal relics, the UV freeze-in production of DM~\cite{Arcadi:2024wwg, Mondal:2025awq, Ikemoto:2022qxy, Biswas:2019iqm, Chen:2017kvz, Wang:2022ojc, Ahmed:2022tfm, Becker:2023tvd}, non-thermal leptogenesis \cite{Asaka:1999yd, Asaka:1999jb, Hamaguchi:2002vc, Fukuyama:2005us, Barman:2021tgt, Ghoshal:2022fud, Zhang:2023oyo},  to the best of our knowledge, the impacts of the inflationary observables on DM production have not been discussed in the literature before.  We find that the Planck 2018 data alone can constrain the cut-off scale associated with UV freeze-in. Moreover, this bound can be significantly improved once we consider the combined data from Planck 2018 and
ACT. In the next section, we briefly discuss the connection between $\trh$ and inflationary observables in the context of $\alpha-$attractor model of inflation. 

\section{Connecting inflationary observables with $\trh$}
\label{sec:inflation_and_TRH}
There are only a few single-field models of inflation that are still allowed by current cosmological data. Among them, the $\alpha$-attractor model is characterized by its ability to predict a universal attractor behavior in the inflationary observables, making them robust against the changes in the model details. The general form of the inflaton potential in the $\alpha$-attractor model (also known as E model) is given by \cite{Kallosh:2013maa},
\bal 
V(\Phi)=\Lambda_s^4 \left(1-e^{-\sqrt{\frac{2}{3\alpha}}\frac{\Phi}{M_P}}\right)^{2n}
 \label{eq:InfPot},
\eal
where $\Lambda_s$ represents the energy scale of the inflation and $M_P$ stands for the reduced Planck mass. $\alpha$ determines how steeply the inflaton field rolls whereas $n$ governs the shape of the potential near its minimum. These parameters influence the rate of field evolution and can impact the duration of inflation and reheating. One can mimic the standard Higgs-Starobinsky inflaton potential by choosing $\alpha=1$ and $n=1$ in Eq. \eqref{eq:InfPot} \cite{Bezrukov:2007ep}. The flatness of the potential can be expressed in terms of the {\it slow-roll parameters} as 
\bal
    \epsilon = \frac{1}{2} M_{P}^2 \left(\frac{\partial_{\Phi} V(\Phi)}{V(\Phi)} \right)^2,\, \eta = M_{P}^2 \left(\frac{\partial_{\Phi}^2 V(\Phi)}{V(\Phi)} \right)\,.
    \label{eq:slowroll}
\eal
By equating one of the above {\it slow-roll parameters} to be unity ($\text{max}[\epsilon, \eta]$), one can find out the inflaton field value ($\Phi_{\rm end}$) at the of the inflation. In case of $\alpha-$ attractor inflation, this is given by  
\bal
\Phi_{\rm end} = \sqrt{\frac{3\alpha}{2}} M_{P} \ln\left(\frac{2n}{\sqrt{3\alpha}} +1\right)\,.
    \label{eq:phiend}
\eal
During the inflation the energy density of the inflaton can be written as
\bal
    \rho = \frac{1}{2} \dot{\Phi}^2 + V(\Phi)\,,
\eal
and the equation of motion of the inflaton field can be written as 
\bal
    \ddot{\Phi} + 3H\dot{\Phi} + V^\prime(\Phi) = 0\,,
    \label{eq:EOM}
\eal
where the Hubble parameter is given by
\bal
H \simeq \sqrt{\dfrac{V(\Phi)}{3 M_P^2}}
\label{eq:H}
\eal

Assuming $\ddot{\Phi} \ll 3 H \dot{\Phi}, V^\prime(\Phi)$, one can write 
\bal
    \rho = V(\Phi) \left( 1+ \frac{\epsilon}{3}\right)\,.
\eal
Hence, the energy density at the end of the inflation, defined by $\epsilon=1$, can be expressed as 
\bal
    \rho_{\rm end} \approx \frac{4}{3} V_{\rm end}\,.
    \label{eq:rhoend}
\eal
In this context, the most important CMB observables that are often used to constrain the inflationary model parameters are the amplitude of the scalar perturbations $A_s$, the tensor-to-scalar ratio $r$, and the spectral index $n_s$. These observables are evaluated at some reference scale known as the {\it pivot scale}, i.e., a specific mode of the inflaton fluctuation with a comoving wave number $k$. By calculating the {\it slow-roll parameters} given in Eq.\, \eqref{eq:slowroll} and the value of $H$ when the mode $k$ crosses the horizon, these observables can be written as 
\begin{eqnarray}
n_s = 1-6\epsilon_k + 2 \eta_k,\,\, r = 16\epsilon_k,\,\,
A_s = \frac{2 H_k^2}{ \pi^2 M_P^2 r}
\label{eq:CMBobs}
\end{eqnarray}
By combining Eqs.\, \eqref{eq:InfPot}, \eqref{eq:slowroll}, and \eqref{eq:CMBobs}, one can write the $n_s$ and $r$ in terms of the model parameters and the field value as,
\begin{eqnarray}
    n_s  =  1 - \frac{8n(e^{\zeta_k} + n)}{3 \alpha (e^{\zeta_k} -1)^2},\,
    r = \frac{64 n^2}{3\alpha (e^{\zeta_k} -1)^2}
    \label{nsandr}
\end{eqnarray}
where $\zeta_k = \sqrt{\frac{2}{3 \alpha}} \frac{\Phi_k}{M_P}$. One can express the field value at the horizon exit $\Phi_k$ can be expressed in terms of the $n_s$ by combining Eq.\,\eqref{eq:slowroll} and Eq.\,\eqref{eq:CMBobs} as 
 \begin{eqnarray}
    \Phi_k &=& \sqrt{\frac{3\alpha}{2}} M_P \ln\left( 1+\right. \nonumber \\
    && \left.\frac{4n + \sqrt{16n^2 + 24 \alpha n (1-n_s)(1+n)}}{3\alpha (1-n_s)}\right).
    \label{eq:phik}
\end{eqnarray} 
Interestingly, by combining both the equations given in \eqref{nsandr}, one can also express $r$ as a function of $n_s$ and the model parameters by combining the above two equations 
\bal
    r = \frac{192 \alpha n^2 (1-n_s)^2}{[4n + \sqrt{16 n^2 + 24 \alpha n (1-n_s)(1+n)}]^2}
\eal
Finally the parameter $\Lambda_s$ can be found by plugging Eq. \eqref{eq:phik} into Eq. \eqref{eq:H} and combining with last equation shown in Eq. \eqref{eq:CMBobs}. We obtain 
\begin{eqnarray}
\Lambda_s &=& M_{\rm P} \left(\frac{3 \pi^2 r A_s}{2} \right)^{1/4} \nonumber \\ && \left[\frac{2n(2n+1)+\sqrt{4n^2 + 6\alpha(1+n) (1-n_s)}}{4n(n+1)} \right]^{n/2}
\end{eqnarray}

Let us now explain how these CMB observables are connected with the reheating dynamics. The horizon exit of a given mode $k$ can be defined as $k = a_k  H_k$ or it can also be written as 
\begin{eqnarray}
    & & \ln\left(\frac{k}{a_k H_k}\right) = \ln\left(\frac{a_{\rm end}}{a_k} \frac{a_{\rm re}}{a_{\rm end}} \frac{a_{0}}{a_{\rm re}} \frac{k}{a_0 H_k}\right) = \nonumber \\
    & & \ln\left(\frac{a_{\rm end}}{a_k}\right) + \ln\left(\frac{a_{\rm re}}{a_{\rm end}}\right) + \ln\left( \frac{a_{0}}{a_{\rm re}}\right) + \ln \left( \frac{k}{a_0 H_k}\right) = 0
    \label{eq:Nknre}
\end{eqnarray}
The above equation can be interpreted in terms of the number of {\it e-folds} $N_k$, the e-folding number from the horizon exit of $k$-mode to the end of inflation and $N_{\rm re}$, the e-folding number from the end of inflation to the end of the reheating, as follows,
\begin{eqnarray}
    N_k + N_{\rm re} + \ln\left( \frac{a_{0}}{a_{\rm re}}\right) + \ln \left( \frac{k}{a_0 H_k}\right) = 0.
    \label{eq:akHk}
\end{eqnarray}
Each term of Eq.\,\eqref{eq:akHk} carries important information and can significantly impact the CMB observables. Let us now discuss these connections in terms of the model parameters. The number of {\it e-folds} between the horizon exit and the end of the inflation $N_k$ can be estimated as
\begin{eqnarray}
   N_k =  \ln \left(\frac{a_{\rm end}}{a_k} \right) &=& \int_{\Phi_{k}}^{\Phi_{\rm end}} \frac{H d\Phi}{\dot{\Phi}} \nonumber \\
   &\approx& - \frac{1}{M_P^2} \int_{\Phi_{k}}^{\Phi_{\rm end}} d\Phi \frac{V(\Phi)}{\partial_{\Phi} V(\Phi)}
    \label{eq:Nkdef}
\end{eqnarray}
By integrating over $\Phi$, one can express $N_k$ as
\begin{equation}
    N_k = \frac{3\alpha}{4n} \left[e^{\zeta_k} - e^{\zeta_{\rm end}} - (\zeta_k - \zeta_{\rm end})\right]
    \label{eq:Nk}
\end{equation}
where $\zeta_{\rm end} = \sqrt{\frac{2}{3 \alpha}} \frac{\Phi_{\rm end}}{M_P}$. The field value at the end of the inflation $\Phi_{\rm end}$ is already defined in Eq.\,\eqref{eq:phiend} whereas the field value at the horizon exit $\Phi_k$ is defined in Eq. \eqref{eq:phik}. The number of {\it e-fold} from the end of the inflation to the end of the reheating can be expressed as 
\begin{eqnarray}
    N_{\rm re} = \ln \left(\frac{a_{\rm re}}{a_{\rm end}} \right) = -\frac{1}{3(1+\overline{\omega}_{\rm re})} \ln\left(\frac{\rho_{\rm re}}{\rho_{\rm end}}\right)\,\,,
    \label{eq:Nre}
\end{eqnarray}
where $\rho_{\rm re}$ and $\rho_{\rm end}$ are the energy densities at the end of reheating and the end of the inflation respectively. $\overline{\omega}_{\rm re}$ is the average equation of state during reheating epoch which is defined as 
\begin{eqnarray}
    \overline{\omega}_{\rm re} = \frac{1}{N_{\rm re}} \int_{0}^{N_{\rm re}} \omega(N_e) dN_e\,\,,
\end{eqnarray}
and can be parameterized as $\overline{\omega}_{\rm re} \approx \frac{n-1}{n+1}$ for a generic inflationary potential $V(\Phi) \propto \Phi^n$ during the oscillatory era as previously shown in \cite{Lozanov:2016hid}. The third term in Eq.\,\eqref{eq:Nknre} can be calculated from the entropy conservation principle between the end of reheating and the present epoch as
\begin{figure}[h!]
    \centering
    \includegraphics[width=0.45\textwidth]{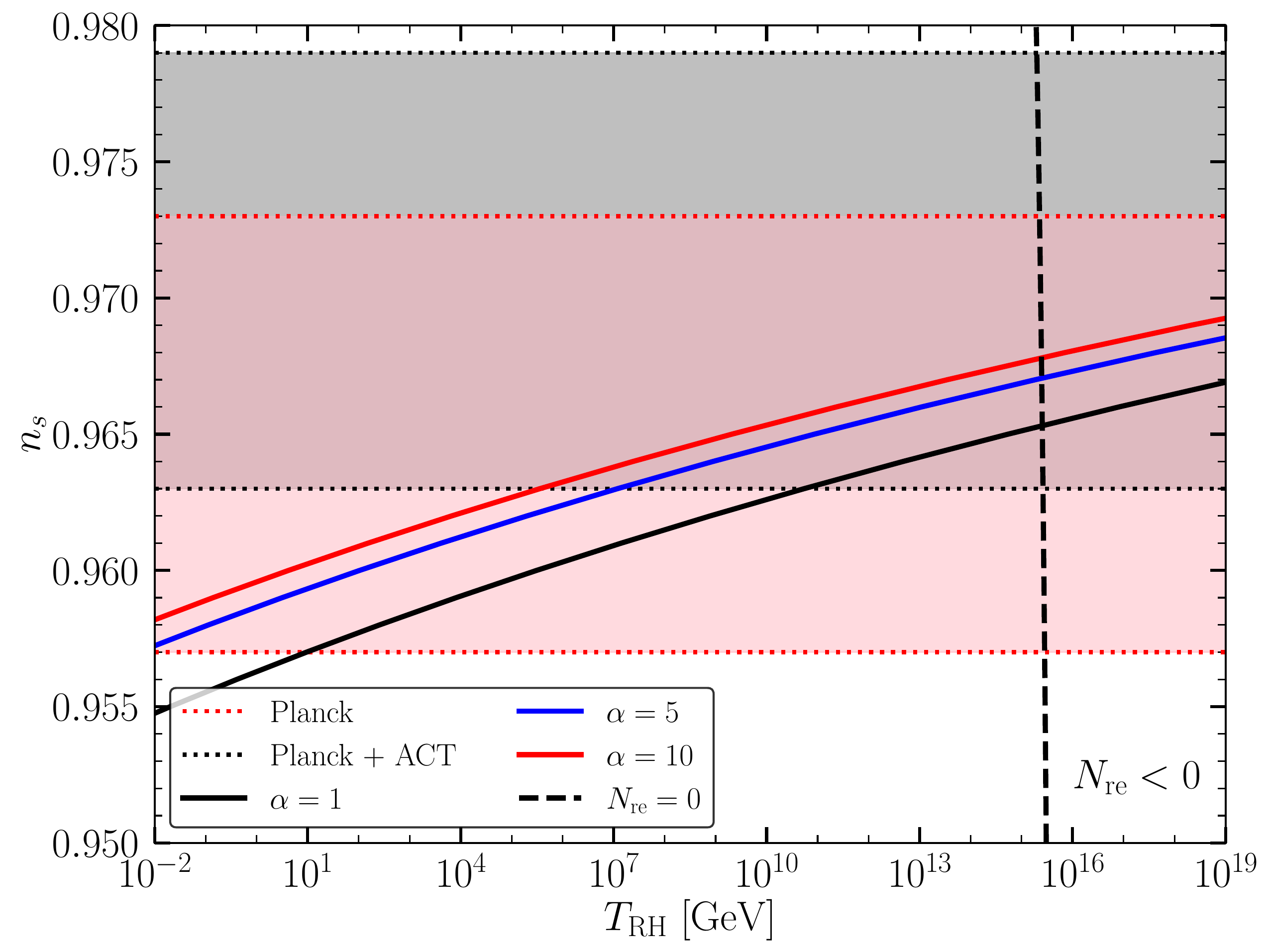}
    \caption{Prediction of $n_s$ as a function of $\trh$  for different choices of the parameter $\alpha$. The red shaded region outlined by red dotted line is the current allowed range of $n_s$ by the Planck 2018 data, whereas the gray region outlined by the black dotted line is the range allowed by Planck and ACT combined.}
    \label{fig:ns_TRH}
\end{figure}
\begin{eqnarray}
    \ln\left(\frac{a_0}{a_{\rm re}}\right) &=& \frac{1}{3}\ln\left(\frac{11 g_{*s}}{43}\right) + \ln \left(\frac{T_{\rm RH}}{T_0} \right) \nonumber \\
&=& \frac{1}{3}\ln\left(\frac{11 g_{*s}}{43}\right) + \frac{1}{4}\ln \left(\frac{30 \rho_{\rm re}}{g_{*} \pi^2 T_0^4} \right) \nonumber \\
&=& \frac{1}{3}\ln\left(\frac{11 g_{*s}}{43}\right) + \frac{1}{4}\ln \left(\frac{30 \rho_{\rm end}}{g_{*} \pi^2 T_0^4}\right)  \nonumber \\
&& - \frac{3}{4}(1+\overline{\omega}_{re}) N_{\rm re}
\label{eq:entropy}
\end{eqnarray}
where in the last line we have replaced $\rho_{\rm re}$ with $\rho_{\rm end}$ and $N_{\rm re}$ from Eq.\,\eqref{eq:Nre}. By using the definition of $H_k$ and $\rho_{\rm end}$ given in Eq.\,\eqref{eq:CMBobs} and \eqref{eq:rhoend}, Eq.\,\eqref{eq:akHk} can now be rewritten as 
\begin{eqnarray}
 N_{\rm re} &=&  \frac{4}{3 \overline{\omega}_{re}-1}\left[N_{k}  + \frac{1}{3}\ln\left(\frac{11 g_{*s}}{43}\right) + \frac{1}{4}\ln \left(\frac{40 }{g_{*} \pi^2}\right)\right.  \nonumber \\ && \left.+ \ln \left(\frac{k}{a_0 T_0}\right) + \frac{1}{2}\ln\left(\frac{2 \sqrt{V_{\rm end}}}{\pi^2 M_P^2 r A_s}\right)\right]\,\,.
 \label{eq:NreRHS}
\end{eqnarray}
During the numerical analysis, we use the value of the pivot scale $k/a_{0}=0.05\text{ Mpc}^{-1}$ where $a_0$ is the scale factor at the present time. From the Planck data \cite{Planck:2018jri}, the observed value of $A_s = 2.099\times10^{-9}$ and the present temperature $T_0 = 2.7 \,\text{K}$. It is crucial to note that the right hand side of the above equation depends only on the parameters of the inflationary model and the CMB observables. On the other hand, by definition given in Eq.\,\eqref{eq:Nre}, the left hand side depends on both the model parameters as well as the coupling between the inflaton and the thermal bath. For a given interaction between the inflaton and the bath particles as well as the inflationary model parameters $\alpha, \text{ and } n$, it is straight forward to calculate the $\rho_{\rm re}$ or in other words the reheating temperature $\trh$. The detailed calculation of $\rho_{\rm re}$ is discussed in the next section. Then, one can equate Eq.\,\eqref{eq:Nre} and \eqref{eq:NreRHS} to find out the corresponding value of the observables $n_s$ and $r$. 
\begin{figure}[h!]
	\centering
	\includegraphics[width = 0.45\textwidth]{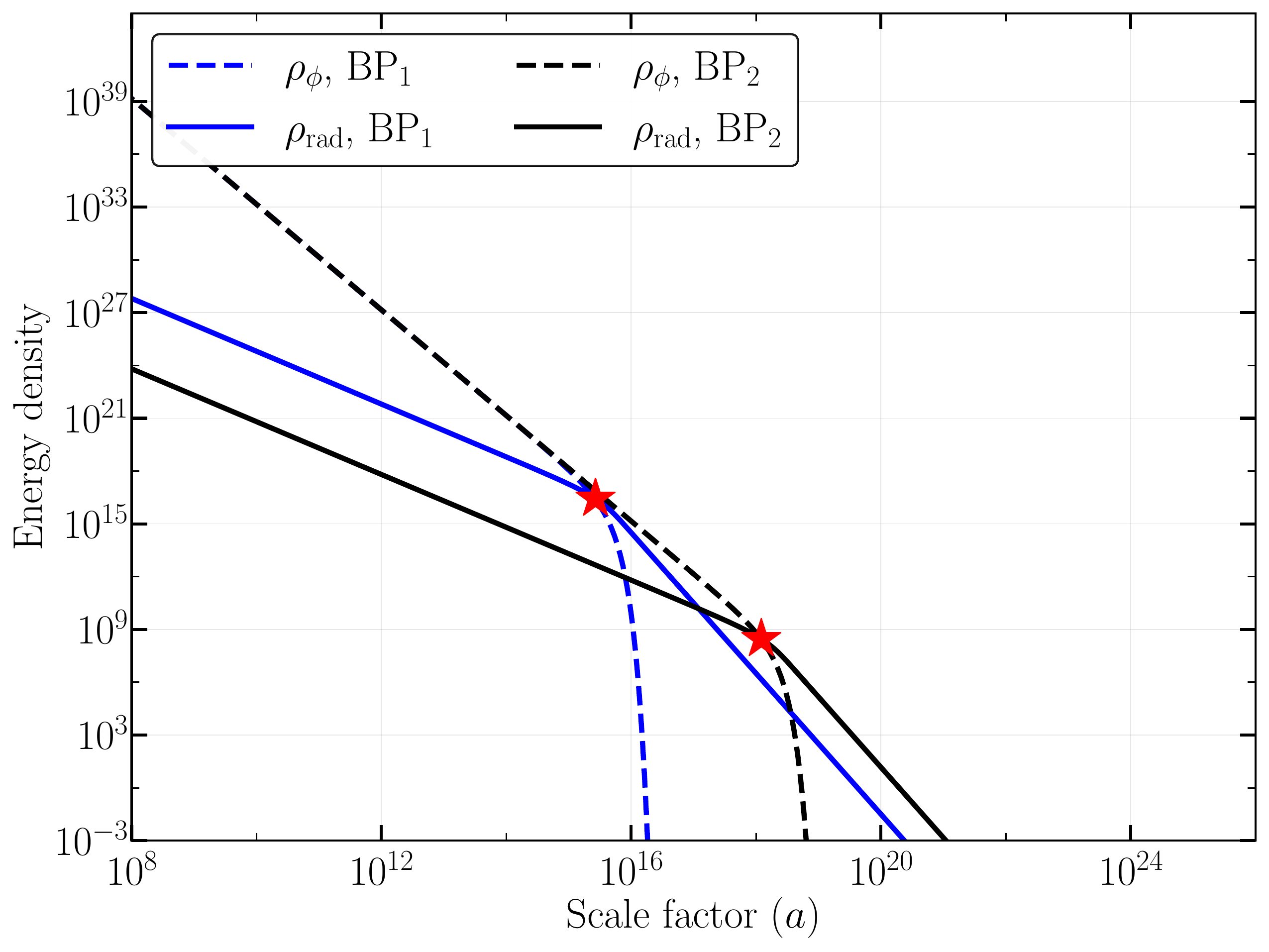} 
        \includegraphics[width = 0.45\textwidth]{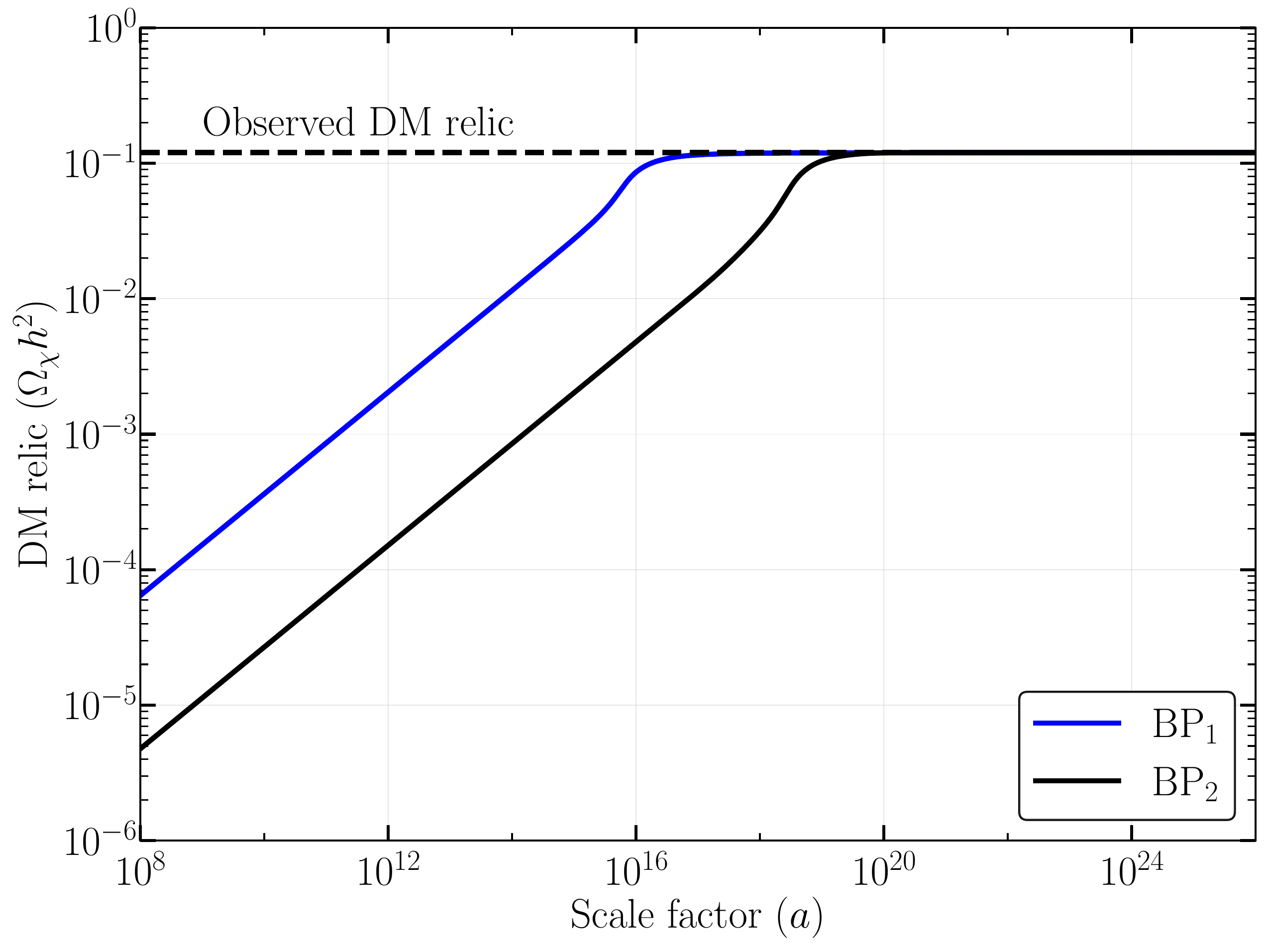}
	\caption{The evolution of energy densities of $\rho_{\Phi}$, $\rho_R$, and the DM relic density $\Omega_{\chi} h^2$ are shown as a function of the scale factor for two different benchmark points (BPs). In both of these BPs, we fix $m_\chi = 1\,\rm GeV$ and consider $(\Gamma_\Phi, \,\Lambda) = (10^{-10}\,\rm GeV, \,2.75 \times 10^{13}\,\rm GeV) \equiv \rm BP_1$ and $(10^{-14}\,\rm GeV, \,3.2 \times 10^{12}\,\rm GeV)\equiv \rm BP_2$ to satisfy the correct DM relic density. The red $\star$ denotes the point of reheating for both the BPs.
}
    \label{fig:BE}
\end{figure}
As mentioned earlier, in recent years the prediction of $n_s$ has become very precise. In Fig. \ref{fig:ns_TRH}, we show the allowed range of $n_s$ from the Planck 2018 data with red shaded region, and from the combined Planck and ACT data with gray shaded region. In the same figure, we present the prediction of $n_s$ as a function of $T_{\rm RH}$ for different benchmark values of the parameter $\alpha$. One can see that a larger $T_{\rm RH}$ corresponds to a larger value of $n_s$. In other words, a higher $T_{\rm RH}$ implies a shorter reheating period (i.e., smaller $N_{\rm re}$) which shifts the horizon exit of observable modes to larger inflaton field values $\Phi_k$. Note that the combination of ACT and Planck data imposes a much stronger limit on $T_{\rm RH}$ compared to the Planck data alone. For example, for $n=1$ and $\alpha=1$, the Planck+ACT data can exclude $T_{\rm RH} \lesssim 10^{11} $ GeV, which is significantly stronger than the Planck 2018 constraint on $\trh \lesssim 10\,\rm GeV$. For larger values of $\alpha$, these constraints become weaker. However, Planck+ACT can still exclude the $T_{\rm RH}\lesssim 10^{6}$ GeV for $\alpha = 10$. The black-dashed vertical line corresponds to the instantaneous reheating or $N_{\rm re}=0$ which determines the maximum value of the $T_{\rm RH}\approx 10^{15}$ GeV allowed in $\alpha-$attractor inflation. 

Since reheating sets the stage for the subsequent phenomenon in the early universe, the value of $\trh$ not only influences the CMB observables but also affects various cosmological dynamics. In particular, the temperature of the thermal plasma is also an extremely important quantity for the DM production through UV freeze-in. Hence, one can expect that there should be a correlation between the CMB observables and the energy scale of the DM production. In the next sections, we will first discuss the DM production mechanism in detail and then show that the above mentioned correlation can be used to put significant constraints on the cut-off scale of UV freeze-in. 
\section{Constraints on UV freeze-in dark matter}
\label{subsec:DMproduction}
After the slow-roll phase ends, the inflaton field oscillates around the minimum of its potential and dissipates its energy into the SM particles. These particles subsequently interact among themselves and reach thermal equilibrium at a temperature $T$. We consider a scenario in which DM is produced from this thermal plasma via a dimension-five operator, given by
\bal
{\cal L} \supset \dfrac{\Phi^\dagger \Phi \bar{\chi} \chi}{\Lambda},
\label{eq:lagrangian}
\eal
where $\Phi$ is the SM Higgs doublet and $\chi$ is a SM singlet Dirac fermion serving as the DM candidate. The operator is suppressed by a heavy mass scale $\Lambda$, making the coupling between DM and the SM sufficiently feeble to evade all direct detection constraints. In the presence of this interaction, DM is produced via the process $\phi_i^\dagger \phi_i \to \bar{\chi} \chi$, where $\phi_i$ denotes the complex scalar components of the Higgs doublet, with $i = 1, 2$. The cross-section of this process is given by
\bal
\sigma_{\phi_i^\dagger\phi_i \to \bar{\chi} \chi}
= \dfrac{1}{8 \pi \Lambda^2} \left(1 - \dfrac{4 m_\chi^2}{\hat{s}}\right)^{3/2}\,\,
\label{eq:xsection}
\eal
where $m_\chi$ is the DM mass and $\hat{s}$ is the Mandelstam variable. Here we assume that the DM production happens much before the Electroweak Symmetry Breaking (EWSB) and thus the SM Higgs is massless. Now, in the presence of the interaction given in 
Eq.\eqref{eq:lagrangian}, we can write the following
Boltzmann equations (BE) in order to investigate the DM production from the SM plasma. The BEs for the DM number
density $n_\chi$, inflaton energy density $\rho_\Phi$,
and radiation density $\rho_R$ can be written as
\bal
&\dfrac{d n_\chi}{dt} + 3 H n_\chi =  {\rm Br}_\chi \dfrac{\Gamma_\Phi}{m_\Phi} \rho_\Phi + 
2\left(\dfrac{\delta n_\chi}{\delta t}\right)_{\rm SM \to\chi}\,\,, \label{eq:DM_BE}\\
&\dfrac{d \rho_\Phi}{dt} + 3 H (1 + \overline{\omega}_{\rm re})\rho_\Phi
 = -\Gamma_{\Phi} \rho_\Phi\,\,,\\
 & \dfrac{ds}{dt} + 3 H s
  = (1 - {\rm Br}_\chi)\dfrac{\Gamma_\Phi \rho_\Phi}{T}
 \label{eq:rad_BE}\,\,,
\eal

where $\Gamma_{\Phi}$ is the decay width of the inflation, and $H$ is the Hubble parameter expressed as
\bal
H \approx \sqrt{\dfrac{1}{3 M_{P}^2} \left(\rho_R + \rho_\Phi \right)}\,\,.
\eal
The prefactor in the right-hand side of Eq.\,\eqref{eq:DM_BE} is due to the contribution to the DM by $\chi$ and $\bar{\chi}$. The explicit form of the 
collision integral is given by \cite{Gondolo:1990dk}
\bal
\left(\dfrac{\delta n_\chi}{\delta t}\right)_{\rm SM \to\chi}
 = \dfrac{g_{\phi_i}^2 T^6}{64 \pi^5 \Lambda^2}
 \int_{2r}^\infty dx\, x (x^2 - 4r^2)^{3/2} {\rm K}_1 (x) \,\,,
 \label{eq:colli}
\eal


where $r = m_\chi/T$, $x = \sqrt{\hat{s}}/T$, $g_{\phi_i}=2$, and ${\rm K}_1(x)$ is modified Bessel function of first kind.

In Fig.\,\ref{fig:BE}, we have shown the evolution of the energy densities of both inflaton and the radiation for two different benchmark values of $\Gamma_{\Phi}$ which correspond to two different $\trh$. After reheating, DM particles are subsequently generated produced via a dimension five operator, and it is known as UV freeze-in mechanism. In this case,
the DM relic strongly depends on the $\trh$ and the dependence changes with the dimensionality of the higher dimension operator. Hence, for a given DM mass, a smaller $\trh$ requires a stronger DM interaction to yield the correct relic abundance and  it is evident from figure\,\ref{fig:BE}.

\begin{figure}[h!]
    \centering
    \includegraphics[width = 0.45\textwidth]{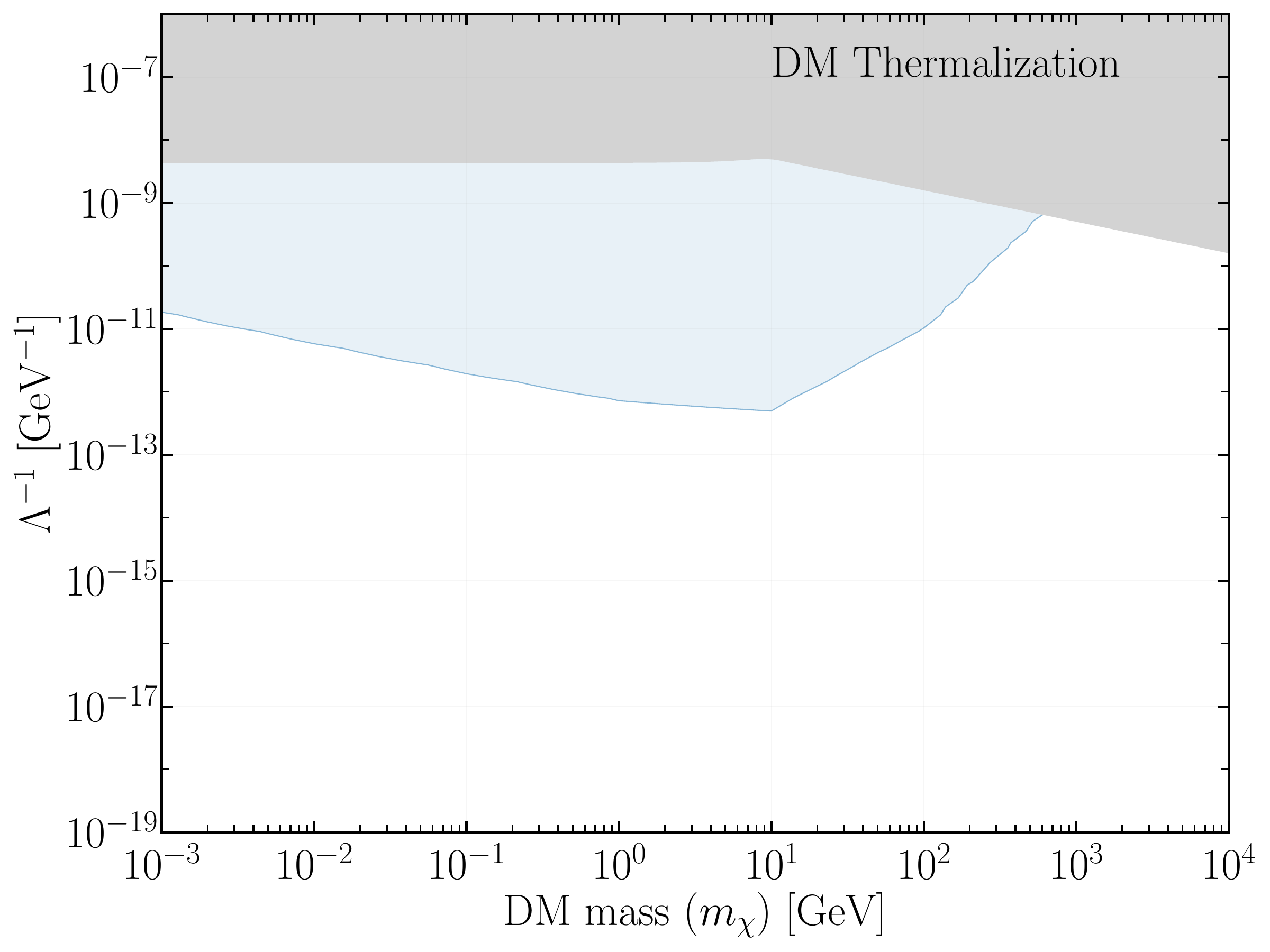}
        \includegraphics[width = 0.45\textwidth]{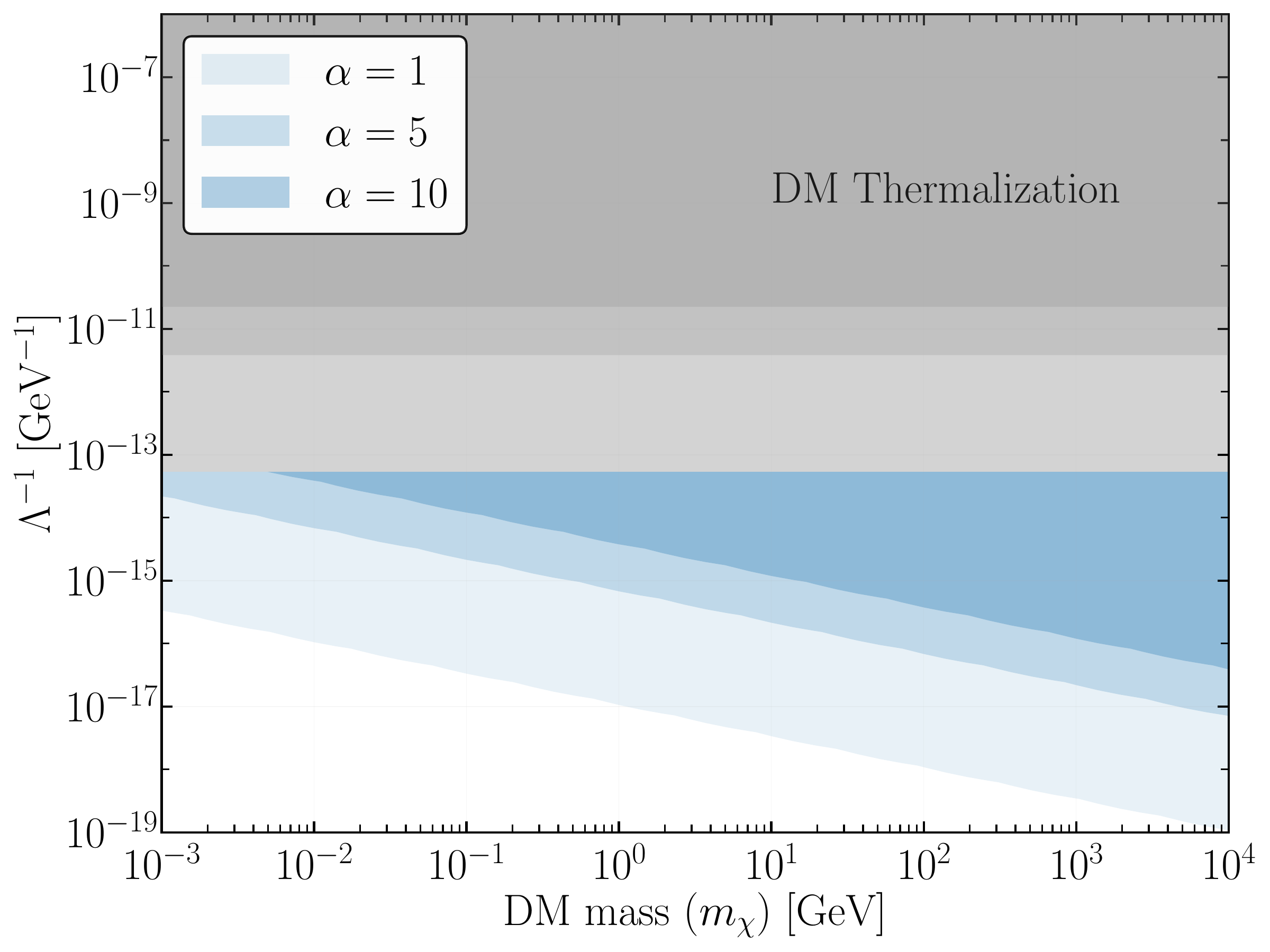}
    \caption{The exclusion limit on the mass scale of the dimension five operator given in Eq.\,\eqref{eq:lagrangian} using Planck 2018 prediction of $n_s$ (upper panel) and Planck+ACT prediction of $n_s$ 
    (lower panel). In the upper panel, we consider $\alpha = 1$. 
    In the lower panel, we consider $\alpha=1,\,5,\,10$. The gray region
    in both panels is disallowed from the DM criterion to be non-thermal. In the lower panel, a darker shade implies a higher value of $\alpha$.}
    \label{fig:smry}
\end{figure}

 Moreover, as discussed in the earlier section that a lower limit on $T_{\rm RH}$ can be derived from a prediction on $n_s$. This lower bound on $T_{\rm RH}$  can therefore be used to constrain DM parameter space. In Fig.\,\ref{fig:smry}, we show the constraint on $m_{\chi} - \Lambda^{-1}$ plane using the $\trh$ prediction by Planck and Planck+ACT data. The shaded blue region in the upper panel is excluded from DM relic density constraint once we use the lower limit on $\trh$, predicted by the Planck 2018 data. In the lower panel, we also show the exclusion limit using Planck+ACT prediction on $\trh$ for different values of $\alpha$. It is important to note that for DM mass between $1\,\rm MeV$ to $1\,\rm TeV$, Planck alone can exclude the cut-off scale $\Lambda \lesssim 10^{12}\,\rm GeV$  which is significantly stronger than any other existing constraints on such minimal scenario. The combined prediction form Planck and ACT data can strengthen the exclusion limit up to the Planck scale for TeV-scale DM. The gray region in both panels  is disallowed from the criterion of DM to be non-thermal and this bound is calculated by comparing the reaction rate $\Gamma_{\chi \to \rm SM}$ with the Hubble parameter at DM production temperature $T_{\rm prod}$. Since  in our scenario, DM is dominantly produced either at $\trh$ when $\trh > m_{\chi}$ or at $m_\chi$ when $\trh < m_\chi$, we consider $T_{\rm prod} = {\rm Max}[\trh, m_\chi]$.


\section{Discussion and conclusion}
\label{sec:conclusion}
In this work, we have demonstrated that how the precise predictions of the inflationary observables, specifically  $n_s$ can be used to constrain the production of UV freeze-in DM through their mutual connection with the $\trh$. The primary objective of this paper was to show that the cosmological data can provide valuable constraints on otherwise elusive UV freeze-in scenario due to their tiny interactions with the SM particles. In particular, we focused on scenarios where, following inflaton decay, DM is produced from the SM thermal plasma after reheating through a dimension-five operator suppressed by a heavy mass scale $\Lambda$. Since the collision term responsible for such production of DM is temperature-dependent, a lower $T_{\rm RH}$ suppresses the DM yield. Consequently, to achieve the observed relic abundance for a given DM mass, a smaller $\trh$ necessitates a stronger interaction i.e. smaller value of the cut-off scale $\Lambda$.

We first demonstrated that precise measurements of $n_s$ can indirectly constrain $T_{\rm RH}$ as discussed previously in the literature. This arises from the fact that a larger $T_{\rm RH}$ corresponds to a shorter reheating phase (smaller \(N_{\rm re}\)), shifting the horizon exit of CMB modes to larger field values. Using the latest Planck and ACT data, we obtained a lower bound on $T_{\rm RH}$, which in turn excludes a extremely large region of the UV freeze-in parameter space shown in $m_\chi\text{--}\Lambda^{-1}$ plane. For illustration purpose, we have considered the $\alpha-$attractor model of inflation. However, this formalism can be used to constrain any single field inflationary model. Let us also note that the $\alpha$-attractior inflationary model with minimal choices of model parameters is excluded at 95\% CL once we combine the data from Planck, ACT, and DESI.

Before ending, we note that our analysis has focused on the case $n = 1$, for which $\overline{\omega}_{\rm re} = 0$, corresponding to a matter-like equation of state. For $ n > 1 $, $\overline{\omega}_{\rm re}$ increases, modifying the dynamics of the inflaton field during reheating. In particular, for  $n = 2$, $\overline{\omega}_{\rm re} = 1/3$, causing the denominator in Eq.~\eqref{eq:Nre} to vanish. As a result, one cannot derive predictions for $N_{\rm re}$ or $\trh$ in the same way (see \cite{Cook:2015vqa} for a related discussion), and  $\trh$ cannot be used to constrain DM production. We have also examined the case $n = 3$, where $\overline{\omega}_{\rm re} = 1/2$, and found that the entire range of $\trh$ remains compatible with the observed values of $n_s$. Consequently, in this case as well, DM production cannot be constrained using $n_s$ by the present data. However, the future improvements in the precision of $n_s$ will open up the possibility of testing even such elusive scenarios, making cosmology a powerful probe of the viability of post-inflationary freeze-in production.

\section*{Acknowledgments}
The work of DN is supported by JSPS Grant-in-Aid for JSPS Research Fellows No. 24KF0238.
The work of SG was supported by the IBS under project code IBS-R018-D1.

\newpage
\bibliography{references}

\begin{thebibliography}{69}%
\makeatletter
\providecommand \@ifxundefined [1]{%
 \@ifx{#1\undefined}
}%
\providecommand \@ifnum [1]{%
 \ifnum #1\expandafter \@firstoftwo
 \else \expandafter \@secondoftwo
 \fi
}%
\providecommand \@ifx [1]{%
 \ifx #1\expandafter \@firstoftwo
 \else \expandafter \@secondoftwo
 \fi
}%
\providecommand \natexlab [1]{#1}%
\providecommand \enquote  [1]{``#1''}%
\providecommand \bibnamefont  [1]{#1}%
\providecommand \bibfnamefont [1]{#1}%
\providecommand \citenamefont [1]{#1}%
\providecommand \href@noop [0]{\@secondoftwo}%
\providecommand \href [0]{\begingroup \@sanitize@url \@href}%
\providecommand \@href[1]{\@@startlink{#1}\@@href}%
\providecommand \@@href[1]{\endgroup#1\@@endlink}%
\providecommand \@sanitize@url [0]{\catcode `\\12\catcode `\$12\catcode
  `\&12\catcode `\#12\catcode `\^12\catcode `\_12\catcode `\%12\relax}%
\providecommand \@@startlink[1]{}%
\providecommand \@@endlink[0]{}%
\providecommand \url  [0]{\begingroup\@sanitize@url \@url }%
\providecommand \@url [1]{\endgroup\@href {#1}{\urlprefix }}%
\providecommand \urlprefix  [0]{URL }%
\providecommand \Eprint [0]{\href }%
\providecommand \doibase [0]{https://doi.org/}%
\providecommand \selectlanguage [0]{\@gobble}%
\providecommand \bibinfo  [0]{\@secondoftwo}%
\providecommand \bibfield  [0]{\@secondoftwo}%
\providecommand \translation [1]{[#1]}%
\providecommand \BibitemOpen [0]{}%
\providecommand \bibitemStop [0]{}%
\providecommand \bibitemNoStop [0]{.\EOS\space}%
\providecommand \EOS [0]{\spacefactor3000\relax}%
\providecommand \BibitemShut  [1]{\csname bibitem#1\endcsname}%
\let\auto@bib@innerbib\@empty
\bibitem [{\citenamefont {Arcadi}\ \emph {et~al.}(2018)\citenamefont {Arcadi},
  \citenamefont {Dutra}, \citenamefont {Ghosh}, \citenamefont {Lindner},
  \citenamefont {Mambrini}, \citenamefont {Pierre}, \citenamefont {Profumo},\
  and\ \citenamefont {Queiroz}}]{Arcadi:2017kky}%
  \BibitemOpen
  \bibfield  {author} {\bibinfo {author} {\bibfnamefont {G.}~\bibnamefont
  {Arcadi}}, \bibinfo {author} {\bibfnamefont {M.}~\bibnamefont {Dutra}},
  \bibinfo {author} {\bibfnamefont {P.}~\bibnamefont {Ghosh}}, \bibinfo
  {author} {\bibfnamefont {M.}~\bibnamefont {Lindner}}, \bibinfo {author}
  {\bibfnamefont {Y.}~\bibnamefont {Mambrini}}, \bibinfo {author}
  {\bibfnamefont {M.}~\bibnamefont {Pierre}}, \bibinfo {author} {\bibfnamefont
  {S.}~\bibnamefont {Profumo}},\ and\ \bibinfo {author} {\bibfnamefont {F.~S.}\
  \bibnamefont {Queiroz}},\ }\bibfield  {title} {\bibinfo {title} {{The waning
  of the WIMP? A review of models, searches, and constraints}},\ }\href
  {https://doi.org/10.1140/epjc/s10052-018-5662-y} {\bibfield  {journal}
  {\bibinfo  {journal} {Eur. Phys. J. C}\ }\textbf {\bibinfo {volume} {78}},\
  \bibinfo {pages} {203} (\bibinfo {year} {2018})},\ \Eprint
  {https://arxiv.org/abs/1703.07364} {arXiv:1703.07364 [hep-ph]} \BibitemShut
  {NoStop}%
\bibitem [{\citenamefont {Roszkowski}\ \emph {et~al.}(2018)\citenamefont
  {Roszkowski}, \citenamefont {Sessolo},\ and\ \citenamefont
  {Trojanowski}}]{Roszkowski:2017nbc}%
  \BibitemOpen
  \bibfield  {author} {\bibinfo {author} {\bibfnamefont {L.}~\bibnamefont
  {Roszkowski}}, \bibinfo {author} {\bibfnamefont {E.~M.}\ \bibnamefont
  {Sessolo}},\ and\ \bibinfo {author} {\bibfnamefont {S.}~\bibnamefont
  {Trojanowski}},\ }\bibfield  {title} {\bibinfo {title} {{WIMP dark matter
  candidates and searches\textemdash{}current status and future prospects}},\
  }\href {https://doi.org/10.1088/1361-6633/aab913} {\bibfield  {journal}
  {\bibinfo  {journal} {Rept. Prog. Phys.}\ }\textbf {\bibinfo {volume} {81}},\
  \bibinfo {pages} {066201} (\bibinfo {year} {2018})},\ \Eprint
  {https://arxiv.org/abs/1707.06277} {arXiv:1707.06277 [hep-ph]} \BibitemShut
  {NoStop}%
\bibitem [{\citenamefont {Arcadi}\ \emph
  {et~al.}(2024{\natexlab{a}})\citenamefont {Arcadi}, \citenamefont
  {Cabo-Almeida}, \citenamefont {Dutra}, \citenamefont {Ghosh}, \citenamefont
  {Lindner}, \citenamefont {Mambrini}, \citenamefont {Neto}, \citenamefont
  {Pierre}, \citenamefont {Profumo},\ and\ \citenamefont
  {Queiroz}}]{Arcadi:2024ukq}%
  \BibitemOpen
  \bibfield  {author} {\bibinfo {author} {\bibfnamefont {G.}~\bibnamefont
  {Arcadi}}, \bibinfo {author} {\bibfnamefont {D.}~\bibnamefont
  {Cabo-Almeida}}, \bibinfo {author} {\bibfnamefont {M.}~\bibnamefont {Dutra}},
  \bibinfo {author} {\bibfnamefont {P.}~\bibnamefont {Ghosh}}, \bibinfo
  {author} {\bibfnamefont {M.}~\bibnamefont {Lindner}}, \bibinfo {author}
  {\bibfnamefont {Y.}~\bibnamefont {Mambrini}}, \bibinfo {author}
  {\bibfnamefont {J.~P.}\ \bibnamefont {Neto}}, \bibinfo {author}
  {\bibfnamefont {M.}~\bibnamefont {Pierre}}, \bibinfo {author} {\bibfnamefont
  {S.}~\bibnamefont {Profumo}},\ and\ \bibinfo {author} {\bibfnamefont {F.~S.}\
  \bibnamefont {Queiroz}},\ }\bibfield  {title} {\bibinfo {title} {{The Waning
  of the WIMP: Endgame?}},\ }\href@noop {} {\  (\bibinfo {year}
  {2024}{\natexlab{a}})},\ \Eprint {https://arxiv.org/abs/2403.15860}
  {arXiv:2403.15860 [hep-ph]} \BibitemShut {NoStop}%
\bibitem [{\citenamefont {Agnese}\ \emph {et~al.}(2016)\citenamefont {Agnese}
  \emph {et~al.}}]{SuperCDMS:2015eex}%
  \BibitemOpen
  \bibfield  {author} {\bibinfo {author} {\bibfnamefont {R.}~\bibnamefont
  {Agnese}} \emph {et~al.} (\bibinfo {collaboration} {SuperCDMS}),\ }\bibfield
  {title} {\bibinfo {title} {{New Results from the Search for Low-Mass Weakly
  Interacting Massive Particles with the CDMS Low Ionization Threshold
  Experiment}},\ }\href {https://doi.org/10.1103/PhysRevLett.116.071301}
  {\bibfield  {journal} {\bibinfo  {journal} {Phys. Rev. Lett.}\ }\textbf
  {\bibinfo {volume} {116}},\ \bibinfo {pages} {071301} (\bibinfo {year}
  {2016})},\ \Eprint {https://arxiv.org/abs/1509.02448} {arXiv:1509.02448
  [astro-ph.CO]} \BibitemShut {NoStop}%
\bibitem [{\citenamefont {Aprile}\ \emph {et~al.}(2018)\citenamefont {Aprile}
  \emph {et~al.}}]{XENON:2018voc}%
  \BibitemOpen
  \bibfield  {author} {\bibinfo {author} {\bibfnamefont {E.}~\bibnamefont
  {Aprile}} \emph {et~al.} (\bibinfo {collaboration} {XENON}),\ }\bibfield
  {title} {\bibinfo {title} {{Dark Matter Search Results from a One Ton-Year
  Exposure of XENON1T}},\ }\href
  {https://doi.org/10.1103/PhysRevLett.121.111302} {\bibfield  {journal}
  {\bibinfo  {journal} {Phys. Rev. Lett.}\ }\textbf {\bibinfo {volume} {121}},\
  \bibinfo {pages} {111302} (\bibinfo {year} {2018})},\ \Eprint
  {https://arxiv.org/abs/1805.12562} {arXiv:1805.12562 [astro-ph.CO]}
  \BibitemShut {NoStop}%
\bibitem [{\citenamefont {Aalbers}\ \emph {et~al.}(2023)\citenamefont {Aalbers}
  \emph {et~al.}}]{LZ:2022lsv}%
  \BibitemOpen
  \bibfield  {author} {\bibinfo {author} {\bibfnamefont {J.}~\bibnamefont
  {Aalbers}} \emph {et~al.} (\bibinfo {collaboration} {LZ}),\ }\bibfield
  {title} {\bibinfo {title} {{First Dark Matter Search Results from the
  LUX-ZEPLIN (LZ) Experiment}},\ }\href
  {https://doi.org/10.1103/PhysRevLett.131.041002} {\bibfield  {journal}
  {\bibinfo  {journal} {Phys. Rev. Lett.}\ }\textbf {\bibinfo {volume} {131}},\
  \bibinfo {pages} {041002} (\bibinfo {year} {2023})},\ \Eprint
  {https://arxiv.org/abs/2207.03764} {arXiv:2207.03764 [hep-ex]} \BibitemShut
  {NoStop}%
\bibitem [{\citenamefont {Lattaud}(2023)}]{Lattaud:2022jnq}%
  \BibitemOpen
  \bibfield  {author} {\bibinfo {author} {\bibfnamefont {H.}~\bibnamefont
  {Lattaud}} (\bibinfo {collaboration} {EDELWEISS}),\ }\bibfield  {title}
  {\bibinfo {title} {{Sub-GeV dark matter searches with EDELWEISS: New results
  and prospects}},\ }\href {https://doi.org/10.21468/SciPostPhysProc.12.012}
  {\bibfield  {journal} {\bibinfo  {journal} {SciPost Phys. Proc.}\ }\textbf
  {\bibinfo {volume} {12}},\ \bibinfo {pages} {012} (\bibinfo {year} {2023})},\
  \Eprint {https://arxiv.org/abs/2211.04176} {arXiv:2211.04176 [astro-ph.GA]}
  \BibitemShut {NoStop}%
\bibitem [{\citenamefont {Aprile}\ \emph {et~al.}(2023)\citenamefont {Aprile}
  \emph {et~al.}}]{XENON:2023cxc}%
  \BibitemOpen
  \bibfield  {author} {\bibinfo {author} {\bibfnamefont {E.}~\bibnamefont
  {Aprile}} \emph {et~al.} (\bibinfo {collaboration} {XENON}),\ }\bibfield
  {title} {\bibinfo {title} {{First Dark Matter Search with Nuclear Recoils
  from the XENONnT Experiment}},\ }\href
  {https://doi.org/10.1103/PhysRevLett.131.041003} {\bibfield  {journal}
  {\bibinfo  {journal} {Phys. Rev. Lett.}\ }\textbf {\bibinfo {volume} {131}},\
  \bibinfo {pages} {041003} (\bibinfo {year} {2023})},\ \Eprint
  {https://arxiv.org/abs/2303.14729} {arXiv:2303.14729 [hep-ex]} \BibitemShut
  {NoStop}%
\bibitem [{\citenamefont {Ahnen}\ \emph {et~al.}(2016)\citenamefont {Ahnen}
  \emph {et~al.}}]{MAGIC:2016xys}%
  \BibitemOpen
  \bibfield  {author} {\bibinfo {author} {\bibfnamefont {M.~L.}\ \bibnamefont
  {Ahnen}} \emph {et~al.} (\bibinfo {collaboration} {MAGIC, Fermi-LAT}),\
  }\bibfield  {title} {\bibinfo {title} {{Limits to Dark Matter Annihilation
  Cross-Section from a Combined Analysis of MAGIC and Fermi-LAT Observations of
  Dwarf Satellite Galaxies}},\ }\href
  {https://doi.org/10.1088/1475-7516/2016/02/039} {\bibfield  {journal}
  {\bibinfo  {journal} {JCAP}\ }\textbf {\bibinfo {volume} {02}},\ \bibinfo
  {pages} {039}},\ \Eprint {https://arxiv.org/abs/1601.06590} {arXiv:1601.06590
  [astro-ph.HE]} \BibitemShut {NoStop}%
\bibitem [{\citenamefont {Abdalla}\ \emph {et~al.}(2022)\citenamefont {Abdalla}
  \emph {et~al.}}]{HESS:2022ygk}%
  \BibitemOpen
  \bibfield  {author} {\bibinfo {author} {\bibfnamefont {H.}~\bibnamefont
  {Abdalla}} \emph {et~al.} (\bibinfo {collaboration} {H.E.S.S.}),\ }\bibfield
  {title} {\bibinfo {title} {{Search for Dark Matter Annihilation Signals in
  the H.E.S.S. Inner Galaxy Survey}},\ }\href
  {https://doi.org/10.1103/PhysRevLett.129.111101} {\bibfield  {journal}
  {\bibinfo  {journal} {Phys. Rev. Lett.}\ }\textbf {\bibinfo {volume} {129}},\
  \bibinfo {pages} {111101} (\bibinfo {year} {2022})},\ \Eprint
  {https://arxiv.org/abs/2207.10471} {arXiv:2207.10471 [astro-ph.HE]}
  \BibitemShut {NoStop}%
\bibitem [{\citenamefont {Kusenko}(2006)}]{Kusenko:2006rh}%
  \BibitemOpen
  \bibfield  {author} {\bibinfo {author} {\bibfnamefont {A.}~\bibnamefont
  {Kusenko}},\ }\bibfield  {title} {\bibinfo {title} {{Sterile neutrinos, dark
  matter, and the pulsar velocities in models with a Higgs singlet}},\ }\href
  {https://doi.org/10.1103/PhysRevLett.97.241301} {\bibfield  {journal}
  {\bibinfo  {journal} {Phys. Rev. Lett.}\ }\textbf {\bibinfo {volume} {97}},\
  \bibinfo {pages} {241301} (\bibinfo {year} {2006})},\ \Eprint
  {https://arxiv.org/abs/hep-ph/0609081} {arXiv:hep-ph/0609081} \BibitemShut
  {NoStop}%
\bibitem [{\citenamefont {McDonald}\ and\ \citenamefont
  {Sahu}(2009)}]{McDonald:2008ua}%
  \BibitemOpen
  \bibfield  {author} {\bibinfo {author} {\bibfnamefont {J.}~\bibnamefont
  {McDonald}}\ and\ \bibinfo {author} {\bibfnamefont {N.}~\bibnamefont
  {Sahu}},\ }\bibfield  {title} {\bibinfo {title} {{keV Warm Dark Matter via
  the Supersymmetric Higgs Portal}},\ }\href
  {https://doi.org/10.1103/PhysRevD.79.103523} {\bibfield  {journal} {\bibinfo
  {journal} {Phys. Rev. D}\ }\textbf {\bibinfo {volume} {79}},\ \bibinfo
  {pages} {103523} (\bibinfo {year} {2009})},\ \Eprint
  {https://arxiv.org/abs/0809.0247} {arXiv:0809.0247 [hep-ph]} \BibitemShut
  {NoStop}%
\bibitem [{\citenamefont {Hall}\ \emph {et~al.}(2010)\citenamefont {Hall},
  \citenamefont {Jedamzik}, \citenamefont {March-Russell},\ and\ \citenamefont
  {West}}]{Hall:2009bx}%
  \BibitemOpen
  \bibfield  {author} {\bibinfo {author} {\bibfnamefont {L.~J.}\ \bibnamefont
  {Hall}}, \bibinfo {author} {\bibfnamefont {K.}~\bibnamefont {Jedamzik}},
  \bibinfo {author} {\bibfnamefont {J.}~\bibnamefont {March-Russell}},\ and\
  \bibinfo {author} {\bibfnamefont {S.~M.}\ \bibnamefont {West}},\ }\bibfield
  {title} {\bibinfo {title} {{Freeze-In Production of FIMP Dark Matter}},\
  }\href {https://doi.org/10.1007/JHEP03(2010)080} {\bibfield  {journal}
  {\bibinfo  {journal} {JHEP}\ }\textbf {\bibinfo {volume} {03}},\ \bibinfo
  {pages} {080}},\ \Eprint {https://arxiv.org/abs/0911.1120} {arXiv:0911.1120
  [hep-ph]} \BibitemShut {NoStop}%
\bibitem [{\citenamefont {Elahi}\ \emph {et~al.}(2015)\citenamefont {Elahi},
  \citenamefont {Kolda},\ and\ \citenamefont {Unwin}}]{Elahi:2014fsa}%
  \BibitemOpen
  \bibfield  {author} {\bibinfo {author} {\bibfnamefont {F.}~\bibnamefont
  {Elahi}}, \bibinfo {author} {\bibfnamefont {C.}~\bibnamefont {Kolda}},\ and\
  \bibinfo {author} {\bibfnamefont {J.}~\bibnamefont {Unwin}},\ }\bibfield
  {title} {\bibinfo {title} {{UltraViolet Freeze-in}},\ }\href
  {https://doi.org/10.1007/JHEP03(2015)048} {\bibfield  {journal} {\bibinfo
  {journal} {JHEP}\ }\textbf {\bibinfo {volume} {03}},\ \bibinfo {pages}
  {048}},\ \Eprint {https://arxiv.org/abs/1410.6157} {arXiv:1410.6157 [hep-ph]}
  \BibitemShut {NoStop}%
\bibitem [{\citenamefont {Bernal}\ \emph {et~al.}(2017)\citenamefont {Bernal},
  \citenamefont {Heikinheimo}, \citenamefont {Tenkanen}, \citenamefont
  {Tuominen},\ and\ \citenamefont {Vaskonen}}]{Bernal:2017kxu}%
  \BibitemOpen
  \bibfield  {author} {\bibinfo {author} {\bibfnamefont {N.}~\bibnamefont
  {Bernal}}, \bibinfo {author} {\bibfnamefont {M.}~\bibnamefont {Heikinheimo}},
  \bibinfo {author} {\bibfnamefont {T.}~\bibnamefont {Tenkanen}}, \bibinfo
  {author} {\bibfnamefont {K.}~\bibnamefont {Tuominen}},\ and\ \bibinfo
  {author} {\bibfnamefont {V.}~\bibnamefont {Vaskonen}},\ }\bibfield  {title}
  {\bibinfo {title} {{The Dawn of FIMP Dark Matter: A Review of Models and
  Constraints}},\ }\href {https://doi.org/10.1142/S0217751X1730023X} {\bibfield
   {journal} {\bibinfo  {journal} {Int. J. Mod. Phys. A}\ }\textbf {\bibinfo
  {volume} {32}},\ \bibinfo {pages} {1730023} (\bibinfo {year} {2017})},\
  \Eprint {https://arxiv.org/abs/1706.07442} {arXiv:1706.07442 [hep-ph]}
  \BibitemShut {NoStop}%
\bibitem [{\citenamefont {Giudice}\ \emph {et~al.}(2001)\citenamefont
  {Giudice}, \citenamefont {Kolb},\ and\ \citenamefont
  {Riotto}}]{Giudice:2000ex}%
  \BibitemOpen
  \bibfield  {author} {\bibinfo {author} {\bibfnamefont {G.~F.}\ \bibnamefont
  {Giudice}}, \bibinfo {author} {\bibfnamefont {E.~W.}\ \bibnamefont {Kolb}},\
  and\ \bibinfo {author} {\bibfnamefont {A.}~\bibnamefont {Riotto}},\
  }\bibfield  {title} {\bibinfo {title} {{Largest temperature of the radiation
  era and its cosmological implications}},\ }\href
  {https://doi.org/10.1103/PhysRevD.64.023508} {\bibfield  {journal} {\bibinfo
  {journal} {Phys. Rev. D}\ }\textbf {\bibinfo {volume} {64}},\ \bibinfo
  {pages} {023508} (\bibinfo {year} {2001})},\ \Eprint
  {https://arxiv.org/abs/hep-ph/0005123} {arXiv:hep-ph/0005123} \BibitemShut
  {NoStop}%
\bibitem [{\citenamefont {Kofman}\ \emph {et~al.}(1997)\citenamefont {Kofman},
  \citenamefont {Linde},\ and\ \citenamefont {Starobinsky}}]{Kofman:1997yn}%
  \BibitemOpen
  \bibfield  {author} {\bibinfo {author} {\bibfnamefont {L.}~\bibnamefont
  {Kofman}}, \bibinfo {author} {\bibfnamefont {A.~D.}\ \bibnamefont {Linde}},\
  and\ \bibinfo {author} {\bibfnamefont {A.~A.}\ \bibnamefont {Starobinsky}},\
  }\bibfield  {title} {\bibinfo {title} {{Towards the theory of reheating after
  inflation}},\ }\href {https://doi.org/10.1103/PhysRevD.56.3258} {\bibfield
  {journal} {\bibinfo  {journal} {Phys. Rev. D}\ }\textbf {\bibinfo {volume}
  {56}},\ \bibinfo {pages} {3258} (\bibinfo {year} {1997})},\ \Eprint
  {https://arxiv.org/abs/hep-ph/9704452} {arXiv:hep-ph/9704452} \BibitemShut
  {NoStop}%
\bibitem [{\citenamefont {Ichikawa}\ \emph {et~al.}(2005)\citenamefont
  {Ichikawa}, \citenamefont {Kawasaki},\ and\ \citenamefont
  {Takahashi}}]{Ichikawa:2005vw}%
  \BibitemOpen
  \bibfield  {author} {\bibinfo {author} {\bibfnamefont {K.}~\bibnamefont
  {Ichikawa}}, \bibinfo {author} {\bibfnamefont {M.}~\bibnamefont {Kawasaki}},\
  and\ \bibinfo {author} {\bibfnamefont {F.}~\bibnamefont {Takahashi}},\
  }\bibfield  {title} {\bibinfo {title} {{The Oscillation effects on
  thermalization of the neutrinos in the Universe with low reheating
  temperature}},\ }\href {https://doi.org/10.1103/PhysRevD.72.043522}
  {\bibfield  {journal} {\bibinfo  {journal} {Phys. Rev. D}\ }\textbf {\bibinfo
  {volume} {72}},\ \bibinfo {pages} {043522} (\bibinfo {year} {2005})},\
  \Eprint {https://arxiv.org/abs/astro-ph/0505395} {arXiv:astro-ph/0505395}
  \BibitemShut {NoStop}%
\bibitem [{\citenamefont {Kawasaki}\ \emph {et~al.}(2000)\citenamefont
  {Kawasaki}, \citenamefont {Kohri},\ and\ \citenamefont
  {Sugiyama}}]{Kawasaki:2000en}%
  \BibitemOpen
  \bibfield  {author} {\bibinfo {author} {\bibfnamefont {M.}~\bibnamefont
  {Kawasaki}}, \bibinfo {author} {\bibfnamefont {K.}~\bibnamefont {Kohri}},\
  and\ \bibinfo {author} {\bibfnamefont {N.}~\bibnamefont {Sugiyama}},\
  }\bibfield  {title} {\bibinfo {title} {{MeV scale reheating temperature and
  thermalization of neutrino background}},\ }\href
  {https://doi.org/10.1103/PhysRevD.62.023506} {\bibfield  {journal} {\bibinfo
  {journal} {Phys. Rev. D}\ }\textbf {\bibinfo {volume} {62}},\ \bibinfo
  {pages} {023506} (\bibinfo {year} {2000})},\ \Eprint
  {https://arxiv.org/abs/astro-ph/0002127} {arXiv:astro-ph/0002127}
  \BibitemShut {NoStop}%
\bibitem [{\citenamefont {Barbieri}\ \emph {et~al.}(2025)\citenamefont
  {Barbieri}, \citenamefont {Brinckmann}, \citenamefont {Gariazzo},
  \citenamefont {Lattanzi}, \citenamefont {Pastor},\ and\ \citenamefont
  {Pisanti}}]{Barbieri:2025moq}%
  \BibitemOpen
  \bibfield  {author} {\bibinfo {author} {\bibfnamefont {N.}~\bibnamefont
  {Barbieri}}, \bibinfo {author} {\bibfnamefont {T.}~\bibnamefont
  {Brinckmann}}, \bibinfo {author} {\bibfnamefont {S.}~\bibnamefont
  {Gariazzo}}, \bibinfo {author} {\bibfnamefont {M.}~\bibnamefont {Lattanzi}},
  \bibinfo {author} {\bibfnamefont {S.}~\bibnamefont {Pastor}},\ and\ \bibinfo
  {author} {\bibfnamefont {O.}~\bibnamefont {Pisanti}},\ }\bibfield  {title}
  {\bibinfo {title} {{Current constraints on cosmological scenarios with very
  low reheating temperatures}},\ }\href@noop {} {\  (\bibinfo {year} {2025})},\
  \Eprint {https://arxiv.org/abs/2501.01369} {arXiv:2501.01369 [astro-ph.CO]}
  \BibitemShut {NoStop}%
\bibitem [{\citenamefont {Baumann}(2011)}]{Baumann:2009ds}%
  \BibitemOpen
  \bibfield  {author} {\bibinfo {author} {\bibfnamefont {D.}~\bibnamefont
  {Baumann}},\ }\bibfield  {title} {\bibinfo {title} {{Inflation}},\ }in\ \href
  {https://doi.org/10.1142/9789814327183_0010} {\emph {\bibinfo {booktitle}
  {{Theoretical Advanced Study Institute in Elementary Particle Physics}:
  {Physics of the Large and the Small}}}}\ (\bibinfo {year} {2011})\ pp.\
  \bibinfo {pages} {523--686},\ \Eprint {https://arxiv.org/abs/0907.5424}
  {arXiv:0907.5424 [hep-th]} \BibitemShut {NoStop}%
\bibitem [{\citenamefont {Aghanim}\ \emph {et~al.}(2020)\citenamefont {Aghanim}
  \emph {et~al.}}]{Planck:2018vyg}%
  \BibitemOpen
  \bibfield  {author} {\bibinfo {author} {\bibfnamefont {N.}~\bibnamefont
  {Aghanim}} \emph {et~al.} (\bibinfo {collaboration} {Planck}),\ }\bibfield
  {title} {\bibinfo {title} {{Planck 2018 results. VI. Cosmological
  parameters}},\ }\href {https://doi.org/10.1051/0004-6361/201833910}
  {\bibfield  {journal} {\bibinfo  {journal} {Astron. Astrophys.}\ }\textbf
  {\bibinfo {volume} {641}},\ \bibinfo {pages} {A6} (\bibinfo {year} {2020})},\
  \bibinfo {note} {[Erratum: Astron.Astrophys. 652, C4 (2021)]},\ \Eprint
  {https://arxiv.org/abs/1807.06209} {arXiv:1807.06209 [astro-ph.CO]}
  \BibitemShut {NoStop}%
\bibitem [{\citenamefont {Cook}\ \emph {et~al.}(2015)\citenamefont {Cook},
  \citenamefont {Dimastrogiovanni}, \citenamefont {Easson},\ and\ \citenamefont
  {Krauss}}]{Cook:2015vqa}%
  \BibitemOpen
  \bibfield  {author} {\bibinfo {author} {\bibfnamefont {J.~L.}\ \bibnamefont
  {Cook}}, \bibinfo {author} {\bibfnamefont {E.}~\bibnamefont
  {Dimastrogiovanni}}, \bibinfo {author} {\bibfnamefont {D.~A.}\ \bibnamefont
  {Easson}},\ and\ \bibinfo {author} {\bibfnamefont {L.~M.}\ \bibnamefont
  {Krauss}},\ }\bibfield  {title} {\bibinfo {title} {{Reheating predictions in
  single field inflation}},\ }\href
  {https://doi.org/10.1088/1475-7516/2015/04/047} {\bibfield  {journal}
  {\bibinfo  {journal} {JCAP}\ }\textbf {\bibinfo {volume} {04}},\ \bibinfo
  {pages} {047}},\ \Eprint {https://arxiv.org/abs/1502.04673} {arXiv:1502.04673
  [astro-ph.CO]} \BibitemShut {NoStop}%
\bibitem [{\citenamefont {Ueno}\ and\ \citenamefont
  {Yamamoto}(2016)}]{Ueno:2016dim}%
  \BibitemOpen
  \bibfield  {author} {\bibinfo {author} {\bibfnamefont {Y.}~\bibnamefont
  {Ueno}}\ and\ \bibinfo {author} {\bibfnamefont {K.}~\bibnamefont
  {Yamamoto}},\ }\bibfield  {title} {\bibinfo {title} {{Constraints on
  $\alpha$-attractor inflation and reheating}},\ }\href
  {https://doi.org/10.1103/PhysRevD.93.083524} {\bibfield  {journal} {\bibinfo
  {journal} {Phys. Rev. D}\ }\textbf {\bibinfo {volume} {93}},\ \bibinfo
  {pages} {083524} (\bibinfo {year} {2016})},\ \Eprint
  {https://arxiv.org/abs/1602.07427} {arXiv:1602.07427 [astro-ph.CO]}
  \BibitemShut {NoStop}%
\bibitem [{\citenamefont {Drewes}\ \emph {et~al.}(2017)\citenamefont {Drewes},
  \citenamefont {Kang},\ and\ \citenamefont {Mun}}]{Drewes:2017fmn}%
  \BibitemOpen
  \bibfield  {author} {\bibinfo {author} {\bibfnamefont {M.}~\bibnamefont
  {Drewes}}, \bibinfo {author} {\bibfnamefont {J.~U.}\ \bibnamefont {Kang}},\
  and\ \bibinfo {author} {\bibfnamefont {U.~R.}\ \bibnamefont {Mun}},\
  }\bibfield  {title} {\bibinfo {title} {{CMB constraints on the inflaton
  couplings and reheating temperature in $\alpha$-attractor inflation}},\
  }\href {https://doi.org/10.1007/JHEP11(2017)072} {\bibfield  {journal}
  {\bibinfo  {journal} {JHEP}\ }\textbf {\bibinfo {volume} {11}},\ \bibinfo
  {pages} {072}},\ \Eprint {https://arxiv.org/abs/1708.01197} {arXiv:1708.01197
  [astro-ph.CO]} \BibitemShut {NoStop}%
\bibitem [{\citenamefont {Maity}\ and\ \citenamefont
  {Saha}(2018)}]{Maity:2018dgy}%
  \BibitemOpen
  \bibfield  {author} {\bibinfo {author} {\bibfnamefont {D.}~\bibnamefont
  {Maity}}\ and\ \bibinfo {author} {\bibfnamefont {P.}~\bibnamefont {Saha}},\
  }\bibfield  {title} {\bibinfo {title} {{Connecting CMB anisotropy and cold
  dark matter phenomenology via reheating}},\ }\href
  {https://doi.org/10.1103/PhysRevD.98.103525} {\bibfield  {journal} {\bibinfo
  {journal} {Phys. Rev. D}\ }\textbf {\bibinfo {volume} {98}},\ \bibinfo
  {pages} {103525} (\bibinfo {year} {2018})},\ \Eprint
  {https://arxiv.org/abs/1801.03059} {arXiv:1801.03059 [hep-ph]} \BibitemShut
  {NoStop}%
\bibitem [{\citenamefont {Maity}\ and\ \citenamefont
  {Saha}(2019)}]{Maity:2018exj}%
  \BibitemOpen
  \bibfield  {author} {\bibinfo {author} {\bibfnamefont {D.}~\bibnamefont
  {Maity}}\ and\ \bibinfo {author} {\bibfnamefont {P.}~\bibnamefont {Saha}},\
  }\bibfield  {title} {\bibinfo {title} {{CMB constraints on dark matter
  phenomenology via reheating in Minimal plateau inflation}},\ }\href
  {https://doi.org/10.1016/j.dark.2019.100317} {\bibfield  {journal} {\bibinfo
  {journal} {Phys. Dark Univ.}\ }\textbf {\bibinfo {volume} {25}},\ \bibinfo
  {pages} {100317} (\bibinfo {year} {2019})},\ \Eprint
  {https://arxiv.org/abs/1804.10115} {arXiv:1804.10115 [hep-ph]} \BibitemShut
  {NoStop}%
\bibitem [{\citenamefont {Drewes}(2022)}]{Drewes:2019rxn}%
  \BibitemOpen
  \bibfield  {author} {\bibinfo {author} {\bibfnamefont {M.}~\bibnamefont
  {Drewes}},\ }\bibfield  {title} {\bibinfo {title} {{Measuring the inflaton
  coupling in~the~CMB}},\ }\href
  {https://doi.org/10.1088/1475-7516/2022/09/069} {\bibfield  {journal}
  {\bibinfo  {journal} {JCAP}\ }\textbf {\bibinfo {volume} {09}},\ \bibinfo
  {pages} {069}},\ \Eprint {https://arxiv.org/abs/1903.09599} {arXiv:1903.09599
  [astro-ph.CO]} \BibitemShut {NoStop}%
\bibitem [{\citenamefont {Haque}\ \emph {et~al.}(2020)\citenamefont {Haque},
  \citenamefont {Maity},\ and\ \citenamefont {Saha}}]{Haque:2020zco}%
  \BibitemOpen
  \bibfield  {author} {\bibinfo {author} {\bibfnamefont {M.~R.}\ \bibnamefont
  {Haque}}, \bibinfo {author} {\bibfnamefont {D.}~\bibnamefont {Maity}},\ and\
  \bibinfo {author} {\bibfnamefont {P.}~\bibnamefont {Saha}},\ }\bibfield
  {title} {\bibinfo {title} {{Two-phase reheating: CMB constraints on inflation
  and dark matter phenomenology}},\ }\href
  {https://doi.org/10.1103/PhysRevD.102.083534} {\bibfield  {journal} {\bibinfo
   {journal} {Phys. Rev. D}\ }\textbf {\bibinfo {volume} {102}},\ \bibinfo
  {pages} {083534} (\bibinfo {year} {2020})},\ \Eprint
  {https://arxiv.org/abs/2009.02794} {arXiv:2009.02794 [hep-th]} \BibitemShut
  {NoStop}%
\bibitem [{\citenamefont {Di~Marco}\ and\ \citenamefont
  {Pradisi}(2021)}]{DiMarco:2021xzk}%
  \BibitemOpen
  \bibfield  {author} {\bibinfo {author} {\bibfnamefont {A.}~\bibnamefont
  {Di~Marco}}\ and\ \bibinfo {author} {\bibfnamefont {G.}~\bibnamefont
  {Pradisi}},\ }\bibfield  {title} {\bibinfo {title} {{Variable inflaton
  equation-of-state and reheating}},\ }\href
  {https://doi.org/10.1142/S0217751X21500950} {\bibfield  {journal} {\bibinfo
  {journal} {Int. J. Mod. Phys. A}\ }\textbf {\bibinfo {volume} {36}},\
  \bibinfo {pages} {2150095} (\bibinfo {year} {2021})},\ \Eprint
  {https://arxiv.org/abs/2102.00326} {arXiv:2102.00326 [gr-qc]} \BibitemShut
  {NoStop}%
\bibitem [{\citenamefont {Akrami}\ \emph {et~al.}(2020)\citenamefont {Akrami}
  \emph {et~al.}}]{Planck:2018jri}%
  \BibitemOpen
  \bibfield  {author} {\bibinfo {author} {\bibfnamefont {Y.}~\bibnamefont
  {Akrami}} \emph {et~al.} (\bibinfo {collaboration} {Planck}),\ }\bibfield
  {title} {\bibinfo {title} {{Planck 2018 results. X. Constraints on
  inflation}},\ }\href {https://doi.org/10.1051/0004-6361/201833887} {\bibfield
   {journal} {\bibinfo  {journal} {Astron. Astrophys.}\ }\textbf {\bibinfo
  {volume} {641}},\ \bibinfo {pages} {A10} (\bibinfo {year} {2020})},\ \Eprint
  {https://arxiv.org/abs/1807.06211} {arXiv:1807.06211 [astro-ph.CO]}
  \BibitemShut {NoStop}%
\bibitem [{\citenamefont {Calabrese}\ \emph {et~al.}(2025)\citenamefont
  {Calabrese} \emph {et~al.}}]{ACT:2025tim}%
  \BibitemOpen
  \bibfield  {author} {\bibinfo {author} {\bibfnamefont {E.}~\bibnamefont
  {Calabrese}} \emph {et~al.} (\bibinfo {collaboration} {ACT}),\ }\bibfield
  {title} {\bibinfo {title} {{The Atacama Cosmology Telescope: DR6 Constraints
  on Extended Cosmological Models}},\ }\href@noop {} {\  (\bibinfo {year}
  {2025})},\ \Eprint {https://arxiv.org/abs/2503.14454} {arXiv:2503.14454
  [astro-ph.CO]} \BibitemShut {NoStop}%
\bibitem [{\citenamefont {Louis}\ \emph {et~al.}(2025)\citenamefont {Louis}
  \emph {et~al.}}]{ACT:2025fju}%
  \BibitemOpen
  \bibfield  {author} {\bibinfo {author} {\bibfnamefont {T.}~\bibnamefont
  {Louis}} \emph {et~al.} (\bibinfo {collaboration} {ACT}),\ }\bibfield
  {title} {\bibinfo {title} {{The Atacama Cosmology Telescope: DR6 Power
  Spectra, Likelihoods and $\Lambda$CDM Parameters}},\ }\href@noop {} {\
  (\bibinfo {year} {2025})},\ \Eprint {https://arxiv.org/abs/2503.14452}
  {arXiv:2503.14452 [astro-ph.CO]} \BibitemShut {NoStop}%
\bibitem [{\citenamefont {Kallosh}\ \emph {et~al.}(2025)\citenamefont
  {Kallosh}, \citenamefont {Linde},\ and\ \citenamefont
  {Roest}}]{Kallosh:2025rni}%
  \BibitemOpen
  \bibfield  {author} {\bibinfo {author} {\bibfnamefont {R.}~\bibnamefont
  {Kallosh}}, \bibinfo {author} {\bibfnamefont {A.}~\bibnamefont {Linde}},\
  and\ \bibinfo {author} {\bibfnamefont {D.}~\bibnamefont {Roest}},\ }\bibfield
   {title} {\bibinfo {title} {{A simple scenario for the last ACT}},\
  }\href@noop {} {\  (\bibinfo {year} {2025})},\ \Eprint
  {https://arxiv.org/abs/2503.21030} {arXiv:2503.21030 [hep-th]} \BibitemShut
  {NoStop}%
\bibitem [{\citenamefont {Pan}\ and\ \citenamefont {Ye}(2025)}]{Pan:2025psn}%
  \BibitemOpen
  \bibfield  {author} {\bibinfo {author} {\bibfnamefont {J.}~\bibnamefont
  {Pan}}\ and\ \bibinfo {author} {\bibfnamefont {G.}~\bibnamefont {Ye}},\
  }\bibfield  {title} {\bibinfo {title} {{Non-minimally coupled gravity
  constraints from DESI DR2 data}},\ }\href@noop {} {\  (\bibinfo {year}
  {2025})},\ \Eprint {https://arxiv.org/abs/2503.19898} {arXiv:2503.19898
  [astro-ph.CO]} \BibitemShut {NoStop}%
\bibitem [{\citenamefont {Dioguardi}\ \emph {et~al.}(2025)\citenamefont
  {Dioguardi}, \citenamefont {Iovino},\ and\ \citenamefont
  {Racioppi}}]{Dioguardi:2025vci}%
  \BibitemOpen
  \bibfield  {author} {\bibinfo {author} {\bibfnamefont {C.}~\bibnamefont
  {Dioguardi}}, \bibinfo {author} {\bibfnamefont {A.~J.}\ \bibnamefont
  {Iovino}},\ and\ \bibinfo {author} {\bibfnamefont {A.}~\bibnamefont
  {Racioppi}},\ }\bibfield  {title} {\bibinfo {title} {{Fractional attractors
  in light of the latest ACT observations}},\ }\href@noop {} {\  (\bibinfo
  {year} {2025})},\ \Eprint {https://arxiv.org/abs/2504.02809}
  {arXiv:2504.02809 [gr-qc]} \BibitemShut {NoStop}%
\bibitem [{\citenamefont {Brahma}\ and\ \citenamefont
  {Calder\'on-Figueroa}(2025)}]{Brahma:2025dio}%
  \BibitemOpen
  \bibfield  {author} {\bibinfo {author} {\bibfnamefont {S.}~\bibnamefont
  {Brahma}}\ and\ \bibinfo {author} {\bibfnamefont {J.}~\bibnamefont
  {Calder\'on-Figueroa}},\ }\bibfield  {title} {\bibinfo {title} {{Is the CMB
  revealing signs of pre-inflationary physics?}},\ }\href@noop {} {\  (\bibinfo
  {year} {2025})},\ \Eprint {https://arxiv.org/abs/2504.02746}
  {arXiv:2504.02746 [astro-ph.CO]} \BibitemShut {NoStop}%
\bibitem [{\citenamefont {Gialamas}\ \emph
  {et~al.}(2025{\natexlab{a}})\citenamefont {Gialamas}, \citenamefont {Karam},
  \citenamefont {Racioppi},\ and\ \citenamefont {Raidal}}]{Gialamas:2025kef}%
  \BibitemOpen
  \bibfield  {author} {\bibinfo {author} {\bibfnamefont {I.~D.}\ \bibnamefont
  {Gialamas}}, \bibinfo {author} {\bibfnamefont {A.}~\bibnamefont {Karam}},
  \bibinfo {author} {\bibfnamefont {A.}~\bibnamefont {Racioppi}},\ and\
  \bibinfo {author} {\bibfnamefont {M.}~\bibnamefont {Raidal}},\ }\bibfield
  {title} {\bibinfo {title} {{Has ACT measured radiative corrections to the
  tree-level Higgs-like inflation?}},\ }\href@noop {} {\  (\bibinfo {year}
  {2025}{\natexlab{a}})},\ \Eprint {https://arxiv.org/abs/2504.06002}
  {arXiv:2504.06002 [astro-ph.CO]} \BibitemShut {NoStop}%
\bibitem [{\citenamefont {Antoniadis}\ \emph {et~al.}(2025)\citenamefont
  {Antoniadis}, \citenamefont {Ellis}, \citenamefont {Ke}, \citenamefont
  {Nanopoulos},\ and\ \citenamefont {Olive}}]{Antoniadis:2025pfa}%
  \BibitemOpen
  \bibfield  {author} {\bibinfo {author} {\bibfnamefont {I.}~\bibnamefont
  {Antoniadis}}, \bibinfo {author} {\bibfnamefont {J.}~\bibnamefont {Ellis}},
  \bibinfo {author} {\bibfnamefont {W.}~\bibnamefont {Ke}}, \bibinfo {author}
  {\bibfnamefont {D.~V.}\ \bibnamefont {Nanopoulos}},\ and\ \bibinfo {author}
  {\bibfnamefont {K.~A.}\ \bibnamefont {Olive}},\ }\bibfield  {title} {\bibinfo
  {title} {{How Accidental was Inflation?}},\ }\href@noop {} {\  (\bibinfo
  {year} {2025})},\ \Eprint {https://arxiv.org/abs/2504.12283}
  {arXiv:2504.12283 [hep-ph]} \BibitemShut {NoStop}%
\bibitem [{\citenamefont {Gao}\ \emph {et~al.}(2025)\citenamefont {Gao},
  \citenamefont {Gong}, \citenamefont {Yi},\ and\ \citenamefont
  {Zhang}}]{Gao:2025onc}%
  \BibitemOpen
  \bibfield  {author} {\bibinfo {author} {\bibfnamefont {Q.}~\bibnamefont
  {Gao}}, \bibinfo {author} {\bibfnamefont {Y.}~\bibnamefont {Gong}}, \bibinfo
  {author} {\bibfnamefont {Z.}~\bibnamefont {Yi}},\ and\ \bibinfo {author}
  {\bibfnamefont {F.}~\bibnamefont {Zhang}},\ }\bibfield  {title} {\bibinfo
  {title} {{Non-minimal coupling in light of ACT}},\ }\href@noop {} {\
  (\bibinfo {year} {2025})},\ \Eprint {https://arxiv.org/abs/2504.15218}
  {arXiv:2504.15218 [astro-ph.CO]} \BibitemShut {NoStop}%
\bibitem [{\citenamefont {Wang}(2025)}]{Wang:2025zri}%
  \BibitemOpen
  \bibfield  {author} {\bibinfo {author} {\bibfnamefont {D.}~\bibnamefont
  {Wang}},\ }\bibfield  {title} {\bibinfo {title} {{Evidence for Dynamical Dark
  Matter}},\ }\href@noop {} {\  (\bibinfo {year} {2025})},\ \Eprint
  {https://arxiv.org/abs/2504.21481} {arXiv:2504.21481 [astro-ph.CO]}
  \BibitemShut {NoStop}%
\bibitem [{\citenamefont {Yin}(2025)}]{Yin:2025rrs}%
  \BibitemOpen
  \bibfield  {author} {\bibinfo {author} {\bibfnamefont {W.}~\bibnamefont
  {Yin}},\ }\bibfield  {title} {\bibinfo {title} {{Higgs-like inflation under
  ACTivated mass}},\ }\href@noop {} {\  (\bibinfo {year} {2025})},\ \Eprint
  {https://arxiv.org/abs/2505.03004} {arXiv:2505.03004 [hep-ph]} \BibitemShut
  {NoStop}%
\bibitem [{\citenamefont {Liu}\ \emph {et~al.}(2025)\citenamefont {Liu},
  \citenamefont {Yi},\ and\ \citenamefont {Gong}}]{Liu:2025qca}%
  \BibitemOpen
  \bibfield  {author} {\bibinfo {author} {\bibfnamefont {L.}~\bibnamefont
  {Liu}}, \bibinfo {author} {\bibfnamefont {Z.}~\bibnamefont {Yi}},\ and\
  \bibinfo {author} {\bibfnamefont {Y.}~\bibnamefont {Gong}},\ }\bibfield
  {title} {\bibinfo {title} {{Reconciling Higgs Inflation with ACT Observations
  through Reheating}},\ }\href@noop {} {\  (\bibinfo {year} {2025})},\ \Eprint
  {https://arxiv.org/abs/2505.02407} {arXiv:2505.02407 [astro-ph.CO]}
  \BibitemShut {NoStop}%
\bibitem [{\citenamefont {Gialamas}\ \emph
  {et~al.}(2025{\natexlab{b}})\citenamefont {Gialamas}, \citenamefont
  {Katsoulas},\ and\ \citenamefont {Tamvakis}}]{Gialamas:2025ofz}%
  \BibitemOpen
  \bibfield  {author} {\bibinfo {author} {\bibfnamefont {I.~D.}\ \bibnamefont
  {Gialamas}}, \bibinfo {author} {\bibfnamefont {T.}~\bibnamefont
  {Katsoulas}},\ and\ \bibinfo {author} {\bibfnamefont {K.}~\bibnamefont
  {Tamvakis}},\ }\bibfield  {title} {\bibinfo {title} {{Keeping the relation
  between the Starobinsky model and no-scale supergravity ACTive}},\
  }\href@noop {} {\  (\bibinfo {year} {2025}{\natexlab{b}})},\ \Eprint
  {https://arxiv.org/abs/2505.03608} {arXiv:2505.03608 [gr-qc]} \BibitemShut
  {NoStop}%
\bibitem [{\citenamefont {McDonald}(2025)}]{McDonald:2025odl}%
  \BibitemOpen
  \bibfield  {author} {\bibinfo {author} {\bibfnamefont {J.}~\bibnamefont
  {McDonald}},\ }\bibfield  {title} {\bibinfo {title} {{Higgs Inflation with
  Vector-Like Quark Stabilisation and the ACT spectral index}},\ }\href@noop {}
  {\  (\bibinfo {year} {2025})},\ \Eprint {https://arxiv.org/abs/2505.07488}
  {arXiv:2505.07488 [hep-ph]} \BibitemShut {NoStop}%
\bibitem [{\citenamefont {Kallosh}\ and\ \citenamefont
  {Linde}(2013{\natexlab{a}})}]{Kallosh:2013lkr}%
  \BibitemOpen
  \bibfield  {author} {\bibinfo {author} {\bibfnamefont {R.}~\bibnamefont
  {Kallosh}}\ and\ \bibinfo {author} {\bibfnamefont {A.}~\bibnamefont
  {Linde}},\ }\bibfield  {title} {\bibinfo {title} {{Superconformal
  generalizations of the Starobinsky model}},\ }\href
  {https://doi.org/10.1088/1475-7516/2013/06/028} {\bibfield  {journal}
  {\bibinfo  {journal} {JCAP}\ }\textbf {\bibinfo {volume} {06}},\ \bibinfo
  {pages} {028}},\ \Eprint {https://arxiv.org/abs/1306.3214} {arXiv:1306.3214
  [hep-th]} \BibitemShut {NoStop}%
\bibitem [{\citenamefont {Kallosh}\ and\ \citenamefont
  {Linde}(2013{\natexlab{b}})}]{Kallosh:2013hoa}%
  \BibitemOpen
  \bibfield  {author} {\bibinfo {author} {\bibfnamefont {R.}~\bibnamefont
  {Kallosh}}\ and\ \bibinfo {author} {\bibfnamefont {A.}~\bibnamefont
  {Linde}},\ }\bibfield  {title} {\bibinfo {title} {{Universality Class in
  Conformal Inflation}},\ }\href
  {https://doi.org/10.1088/1475-7516/2013/07/002} {\bibfield  {journal}
  {\bibinfo  {journal} {JCAP}\ }\textbf {\bibinfo {volume} {07}},\ \bibinfo
  {pages} {002}},\ \Eprint {https://arxiv.org/abs/1306.5220} {arXiv:1306.5220
  [hep-th]} \BibitemShut {NoStop}%
\bibitem [{\citenamefont {Kallosh}\ \emph {et~al.}(2013)\citenamefont
  {Kallosh}, \citenamefont {Linde},\ and\ \citenamefont
  {Roest}}]{Kallosh:2013yoa}%
  \BibitemOpen
  \bibfield  {author} {\bibinfo {author} {\bibfnamefont {R.}~\bibnamefont
  {Kallosh}}, \bibinfo {author} {\bibfnamefont {A.}~\bibnamefont {Linde}},\
  and\ \bibinfo {author} {\bibfnamefont {D.}~\bibnamefont {Roest}},\ }\bibfield
   {title} {\bibinfo {title} {{Superconformal Inflationary
  $\alpha$-Attractors}},\ }\href {https://doi.org/10.1007/JHEP11(2013)198}
  {\bibfield  {journal} {\bibinfo  {journal} {JHEP}\ }\textbf {\bibinfo
  {volume} {11}},\ \bibinfo {pages} {198}},\ \Eprint
  {https://arxiv.org/abs/1311.0472} {arXiv:1311.0472 [hep-th]} \BibitemShut
  {NoStop}%
\bibitem [{\citenamefont {Kallosh}\ and\ \citenamefont
  {Linde}(2013{\natexlab{c}})}]{Kallosh:2013pby}%
  \BibitemOpen
  \bibfield  {author} {\bibinfo {author} {\bibfnamefont {R.}~\bibnamefont
  {Kallosh}}\ and\ \bibinfo {author} {\bibfnamefont {A.}~\bibnamefont
  {Linde}},\ }\bibfield  {title} {\bibinfo {title} {{Superconformal
  generalization of the chaotic inflation model $\frac{\lambda}{4} \phi^{4} -
  \frac{\xi}{2} \phi^{2}R$}},\ }\href
  {https://doi.org/10.1088/1475-7516/2013/06/027} {\bibfield  {journal}
  {\bibinfo  {journal} {JCAP}\ }\textbf {\bibinfo {volume} {06}},\ \bibinfo
  {pages} {027}},\ \Eprint {https://arxiv.org/abs/1306.3211} {arXiv:1306.3211
  [hep-th]} \BibitemShut {NoStop}%
\bibitem [{\citenamefont {Kallosh}\ and\ \citenamefont
  {Linde}(2013{\natexlab{d}})}]{Kallosh:2013maa}%
  \BibitemOpen
  \bibfield  {author} {\bibinfo {author} {\bibfnamefont {R.}~\bibnamefont
  {Kallosh}}\ and\ \bibinfo {author} {\bibfnamefont {A.}~\bibnamefont
  {Linde}},\ }\bibfield  {title} {\bibinfo {title} {{Non-minimal Inflationary
  Attractors}},\ }\href {https://doi.org/10.1088/1475-7516/2013/10/033}
  {\bibfield  {journal} {\bibinfo  {journal} {JCAP}\ }\textbf {\bibinfo
  {volume} {10}},\ \bibinfo {pages} {033}},\ \Eprint
  {https://arxiv.org/abs/1307.7938} {arXiv:1307.7938 [hep-th]} \BibitemShut
  {NoStop}%
\bibitem [{\citenamefont {Galante}\ \emph {et~al.}(2015)\citenamefont
  {Galante}, \citenamefont {Kallosh}, \citenamefont {Linde},\ and\
  \citenamefont {Roest}}]{Galante:2014ifa}%
  \BibitemOpen
  \bibfield  {author} {\bibinfo {author} {\bibfnamefont {M.}~\bibnamefont
  {Galante}}, \bibinfo {author} {\bibfnamefont {R.}~\bibnamefont {Kallosh}},
  \bibinfo {author} {\bibfnamefont {A.}~\bibnamefont {Linde}},\ and\ \bibinfo
  {author} {\bibfnamefont {D.}~\bibnamefont {Roest}},\ }\bibfield  {title}
  {\bibinfo {title} {{Unity of Cosmological Inflation Attractors}},\ }\href
  {https://doi.org/10.1103/PhysRevLett.114.141302} {\bibfield  {journal}
  {\bibinfo  {journal} {Phys. Rev. Lett.}\ }\textbf {\bibinfo {volume} {114}},\
  \bibinfo {pages} {141302} (\bibinfo {year} {2015})},\ \Eprint
  {https://arxiv.org/abs/1412.3797} {arXiv:1412.3797 [hep-th]} \BibitemShut
  {NoStop}%
\bibitem [{\citenamefont {Arcadi}\ \emph
  {et~al.}(2024{\natexlab{b}})\citenamefont {Arcadi}, \citenamefont {Costa},
  \citenamefont {Goudelis},\ and\ \citenamefont {Lebedev}}]{Arcadi:2024wwg}%
  \BibitemOpen
  \bibfield  {author} {\bibinfo {author} {\bibfnamefont {G.}~\bibnamefont
  {Arcadi}}, \bibinfo {author} {\bibfnamefont {F.}~\bibnamefont {Costa}},
  \bibinfo {author} {\bibfnamefont {A.}~\bibnamefont {Goudelis}},\ and\
  \bibinfo {author} {\bibfnamefont {O.}~\bibnamefont {Lebedev}},\ }\bibfield
  {title} {\bibinfo {title} {{Higgs portal dark matter freeze-in at stronger
  coupling: observational benchmarks}},\ }\href
  {https://doi.org/10.1007/JHEP07(2024)044} {\bibfield  {journal} {\bibinfo
  {journal} {JHEP}\ }\textbf {\bibinfo {volume} {07}},\ \bibinfo {pages}
  {044}},\ \Eprint {https://arxiv.org/abs/2405.03760} {arXiv:2405.03760
  [hep-ph]} \BibitemShut {NoStop}%
\bibitem [{\citenamefont {Mondal}\ \emph {et~al.}(2025)\citenamefont {Mondal},
  \citenamefont {Mondal},\ and\ \citenamefont {Yamada}}]{Mondal:2025awq}%
  \BibitemOpen
  \bibfield  {author} {\bibinfo {author} {\bibfnamefont {R.}~\bibnamefont
  {Mondal}}, \bibinfo {author} {\bibfnamefont {S.}~\bibnamefont {Mondal}},\
  and\ \bibinfo {author} {\bibfnamefont {T.}~\bibnamefont {Yamada}},\
  }\bibfield  {title} {\bibinfo {title} {{Freeze-in and Freeze-out production
  of Higgs Portal Majorana Fermionic Dark Matter during and after Reheating}},\
  }\href@noop {} {\  (\bibinfo {year} {2025})},\ \Eprint
  {https://arxiv.org/abs/2503.20738} {arXiv:2503.20738 [hep-ph]} \BibitemShut
  {NoStop}%
\bibitem [{\citenamefont {Ikemoto}\ \emph {et~al.}(2023)\citenamefont
  {Ikemoto}, \citenamefont {Haba}, \citenamefont {Yasuhiro},\ and\
  \citenamefont {Yamada}}]{Ikemoto:2022qxy}%
  \BibitemOpen
  \bibfield  {author} {\bibinfo {author} {\bibfnamefont {J.}~\bibnamefont
  {Ikemoto}}, \bibinfo {author} {\bibfnamefont {N.}~\bibnamefont {Haba}},
  \bibinfo {author} {\bibfnamefont {S.}~\bibnamefont {Yasuhiro}},\ and\
  \bibinfo {author} {\bibfnamefont {T.}~\bibnamefont {Yamada}},\ }\bibfield
  {title} {\bibinfo {title} {{Higgs portal majorana fermionic dark matter with
  the freeze-in mechanism}},\ }\href {https://doi.org/10.1093/ptep/ptad081}
  {\bibfield  {journal} {\bibinfo  {journal} {PTEP}\ }\textbf {\bibinfo
  {volume} {2023}},\ \bibinfo {pages} {083B04} (\bibinfo {year} {2023})},\
  \Eprint {https://arxiv.org/abs/2212.14660} {arXiv:2212.14660 [hep-ph]}
  \BibitemShut {NoStop}%
\bibitem [{\citenamefont {Biswas}\ \emph {et~al.}(2020)\citenamefont {Biswas},
  \citenamefont {Ganguly},\ and\ \citenamefont {Roy}}]{Biswas:2019iqm}%
  \BibitemOpen
  \bibfield  {author} {\bibinfo {author} {\bibfnamefont {A.}~\bibnamefont
  {Biswas}}, \bibinfo {author} {\bibfnamefont {S.}~\bibnamefont {Ganguly}},\
  and\ \bibinfo {author} {\bibfnamefont {S.}~\bibnamefont {Roy}},\ }\bibfield
  {title} {\bibinfo {title} {{Fermionic dark matter via UV and IR freeze-in and
  its possible X-ray signature}},\ }\href
  {https://doi.org/10.1088/1475-7516/2020/03/043} {\bibfield  {journal}
  {\bibinfo  {journal} {JCAP}\ }\textbf {\bibinfo {volume} {03}},\ \bibinfo
  {pages} {043}},\ \Eprint {https://arxiv.org/abs/1907.07973} {arXiv:1907.07973
  [hep-ph]} \BibitemShut {NoStop}%
\bibitem [{\citenamefont {Chen}\ and\ \citenamefont
  {Kang}(2018)}]{Chen:2017kvz}%
  \BibitemOpen
  \bibfield  {author} {\bibinfo {author} {\bibfnamefont {S.-L.}\ \bibnamefont
  {Chen}}\ and\ \bibinfo {author} {\bibfnamefont {Z.}~\bibnamefont {Kang}},\
  }\bibfield  {title} {\bibinfo {title} {{On UltraViolet Freeze-in Dark Matter
  during Reheating}},\ }\href {https://doi.org/10.1088/1475-7516/2018/05/036}
  {\bibfield  {journal} {\bibinfo  {journal} {JCAP}\ }\textbf {\bibinfo
  {volume} {05}},\ \bibinfo {pages} {036}},\ \Eprint
  {https://arxiv.org/abs/1711.02556} {arXiv:1711.02556 [hep-ph]} \BibitemShut
  {NoStop}%
\bibitem [{\citenamefont {Wang}\ \emph {et~al.}(2022)\citenamefont {Wang},
  \citenamefont {Tang},\ and\ \citenamefont {Wu}}]{Wang:2022ojc}%
  \BibitemOpen
  \bibfield  {author} {\bibinfo {author} {\bibfnamefont {Q.-Y.}\ \bibnamefont
  {Wang}}, \bibinfo {author} {\bibfnamefont {Y.}~\bibnamefont {Tang}},\ and\
  \bibinfo {author} {\bibfnamefont {Y.-L.}\ \bibnamefont {Wu}},\ }\bibfield
  {title} {\bibinfo {title} {{Dark matter production in Weyl R2 inflation}},\
  }\href {https://doi.org/10.1103/PhysRevD.106.023502} {\bibfield  {journal}
  {\bibinfo  {journal} {Phys. Rev. D}\ }\textbf {\bibinfo {volume} {106}},\
  \bibinfo {pages} {023502} (\bibinfo {year} {2022})},\ \Eprint
  {https://arxiv.org/abs/2203.15452} {arXiv:2203.15452 [hep-ph]} \BibitemShut
  {NoStop}%
\bibitem [{\citenamefont {Ahmed}\ \emph {et~al.}(2023)\citenamefont {Ahmed},
  \citenamefont {Grzadkowski},\ and\ \citenamefont {Socha}}]{Ahmed:2022tfm}%
  \BibitemOpen
  \bibfield  {author} {\bibinfo {author} {\bibfnamefont {A.}~\bibnamefont
  {Ahmed}}, \bibinfo {author} {\bibfnamefont {B.}~\bibnamefont {Grzadkowski}},\
  and\ \bibinfo {author} {\bibfnamefont {A.}~\bibnamefont {Socha}},\ }\bibfield
   {title} {\bibinfo {title} {{Higgs boson induced reheating and ultraviolet
  frozen-in dark matter}},\ }\href {https://doi.org/10.1007/JHEP02(2023)196}
  {\bibfield  {journal} {\bibinfo  {journal} {JHEP}\ }\textbf {\bibinfo
  {volume} {02}},\ \bibinfo {pages} {196}},\ \Eprint
  {https://arxiv.org/abs/2207.11218} {arXiv:2207.11218 [hep-ph]} \BibitemShut
  {NoStop}%
\bibitem [{\citenamefont {Becker}\ \emph {et~al.}(2024)\citenamefont {Becker},
  \citenamefont {Copello}, \citenamefont {Harz}, \citenamefont {Lang},\ and\
  \citenamefont {Xu}}]{Becker:2023tvd}%
  \BibitemOpen
  \bibfield  {author} {\bibinfo {author} {\bibfnamefont {M.}~\bibnamefont
  {Becker}}, \bibinfo {author} {\bibfnamefont {E.}~\bibnamefont {Copello}},
  \bibinfo {author} {\bibfnamefont {J.}~\bibnamefont {Harz}}, \bibinfo {author}
  {\bibfnamefont {J.}~\bibnamefont {Lang}},\ and\ \bibinfo {author}
  {\bibfnamefont {Y.}~\bibnamefont {Xu}},\ }\bibfield  {title} {\bibinfo
  {title} {{Confronting dark matter freeze-in during reheating with constraints
  from inflation}},\ }\href {https://doi.org/10.1088/1475-7516/2024/01/053}
  {\bibfield  {journal} {\bibinfo  {journal} {JCAP}\ }\textbf {\bibinfo
  {volume} {01}},\ \bibinfo {pages} {053}},\ \Eprint
  {https://arxiv.org/abs/2306.17238} {arXiv:2306.17238 [hep-ph]} \BibitemShut
  {NoStop}%
\bibitem [{\citenamefont {Asaka}\ \emph {et~al.}(1999)\citenamefont {Asaka},
  \citenamefont {Hamaguchi}, \citenamefont {Kawasaki},\ and\ \citenamefont
  {Yanagida}}]{Asaka:1999yd}%
  \BibitemOpen
  \bibfield  {author} {\bibinfo {author} {\bibfnamefont {T.}~\bibnamefont
  {Asaka}}, \bibinfo {author} {\bibfnamefont {K.}~\bibnamefont {Hamaguchi}},
  \bibinfo {author} {\bibfnamefont {M.}~\bibnamefont {Kawasaki}},\ and\
  \bibinfo {author} {\bibfnamefont {T.}~\bibnamefont {Yanagida}},\ }\bibfield
  {title} {\bibinfo {title} {{Leptogenesis in inflaton decay}},\ }\href
  {https://doi.org/10.1016/S0370-2693(99)01020-5} {\bibfield  {journal}
  {\bibinfo  {journal} {Phys. Lett. B}\ }\textbf {\bibinfo {volume} {464}},\
  \bibinfo {pages} {12} (\bibinfo {year} {1999})},\ \Eprint
  {https://arxiv.org/abs/hep-ph/9906366} {arXiv:hep-ph/9906366} \BibitemShut
  {NoStop}%
\bibitem [{\citenamefont {Asaka}\ \emph {et~al.}(2000)\citenamefont {Asaka},
  \citenamefont {Hamaguchi}, \citenamefont {Kawasaki},\ and\ \citenamefont
  {Yanagida}}]{Asaka:1999jb}%
  \BibitemOpen
  \bibfield  {author} {\bibinfo {author} {\bibfnamefont {T.}~\bibnamefont
  {Asaka}}, \bibinfo {author} {\bibfnamefont {K.}~\bibnamefont {Hamaguchi}},
  \bibinfo {author} {\bibfnamefont {M.}~\bibnamefont {Kawasaki}},\ and\
  \bibinfo {author} {\bibfnamefont {T.}~\bibnamefont {Yanagida}},\ }\bibfield
  {title} {\bibinfo {title} {{Leptogenesis in inflationary universe}},\ }\href
  {https://doi.org/10.1103/PhysRevD.61.083512} {\bibfield  {journal} {\bibinfo
  {journal} {Phys. Rev. D}\ }\textbf {\bibinfo {volume} {61}},\ \bibinfo
  {pages} {083512} (\bibinfo {year} {2000})},\ \Eprint
  {https://arxiv.org/abs/hep-ph/9907559} {arXiv:hep-ph/9907559} \BibitemShut
  {NoStop}%
\bibitem [{\citenamefont {Hamaguchi}(2002)}]{Hamaguchi:2002vc}%
  \BibitemOpen
  \bibfield  {author} {\bibinfo {author} {\bibfnamefont {K.}~\bibnamefont
  {Hamaguchi}},\ }\emph {\bibinfo {title} {{Cosmological baryon asymmetry and
  neutrinos: Baryogenesis via leptogenesis in supersymmetric theories}}},\
  \href@noop {} {Ph.D. thesis},\ \bibinfo  {school} {Tokyo U.} (\bibinfo {year}
  {2002}),\ \Eprint {https://arxiv.org/abs/hep-ph/0212305}
  {arXiv:hep-ph/0212305} \BibitemShut {NoStop}%
\bibitem [{\citenamefont {Fukuyama}\ \emph {et~al.}(2005)\citenamefont
  {Fukuyama}, \citenamefont {Kikuchi},\ and\ \citenamefont
  {Osaka}}]{Fukuyama:2005us}%
  \BibitemOpen
  \bibfield  {author} {\bibinfo {author} {\bibfnamefont {T.}~\bibnamefont
  {Fukuyama}}, \bibinfo {author} {\bibfnamefont {T.}~\bibnamefont {Kikuchi}},\
  and\ \bibinfo {author} {\bibfnamefont {T.}~\bibnamefont {Osaka}},\ }\bibfield
   {title} {\bibinfo {title} {{Non-thermal leptogenesis and a prediction of
  inflaton mass in a supersymmetric SO(10) model}},\ }\href
  {https://doi.org/10.1088/1475-7516/2005/06/005} {\bibfield  {journal}
  {\bibinfo  {journal} {JCAP}\ }\textbf {\bibinfo {volume} {06}},\ \bibinfo
  {pages} {005}},\ \Eprint {https://arxiv.org/abs/hep-ph/0503201}
  {arXiv:hep-ph/0503201} \BibitemShut {NoStop}%
\bibitem [{\citenamefont {Barman}\ \emph {et~al.}(2021)\citenamefont {Barman},
  \citenamefont {Borah},\ and\ \citenamefont {Roshan}}]{Barman:2021tgt}%
  \BibitemOpen
  \bibfield  {author} {\bibinfo {author} {\bibfnamefont {B.}~\bibnamefont
  {Barman}}, \bibinfo {author} {\bibfnamefont {D.}~\bibnamefont {Borah}},\ and\
  \bibinfo {author} {\bibfnamefont {R.}~\bibnamefont {Roshan}},\ }\bibfield
  {title} {\bibinfo {title} {{Nonthermal leptogenesis and UV freeze-in of dark
  matter: Impact of inflationary reheating}},\ }\href
  {https://doi.org/10.1103/PhysRevD.104.035022} {\bibfield  {journal} {\bibinfo
   {journal} {Phys. Rev. D}\ }\textbf {\bibinfo {volume} {104}},\ \bibinfo
  {pages} {035022} (\bibinfo {year} {2021})},\ \Eprint
  {https://arxiv.org/abs/2103.01675} {arXiv:2103.01675 [hep-ph]} \BibitemShut
  {NoStop}%
\bibitem [{\citenamefont {Ghoshal}\ \emph {et~al.}(2024)\citenamefont
  {Ghoshal}, \citenamefont {Nanda},\ and\ \citenamefont
  {Saha}}]{Ghoshal:2022fud}%
  \BibitemOpen
  \bibfield  {author} {\bibinfo {author} {\bibfnamefont {A.}~\bibnamefont
  {Ghoshal}}, \bibinfo {author} {\bibfnamefont {D.}~\bibnamefont {Nanda}},\
  and\ \bibinfo {author} {\bibfnamefont {A.~K.}\ \bibnamefont {Saha}},\
  }\bibfield  {title} {\bibinfo {title} {{CMB imprints of high scale
  non-thermal leptogenesis}},\ }\href
  {https://doi.org/10.1016/j.physletb.2024.138484} {\bibfield  {journal}
  {\bibinfo  {journal} {Phys. Lett. B}\ }\textbf {\bibinfo {volume} {849}},\
  \bibinfo {pages} {138484} (\bibinfo {year} {2024})},\ \Eprint
  {https://arxiv.org/abs/2210.14176} {arXiv:2210.14176 [hep-ph]} \BibitemShut
  {NoStop}%
\bibitem [{\citenamefont {Zhang}(2024)}]{Zhang:2023oyo}%
  \BibitemOpen
  \bibfield  {author} {\bibinfo {author} {\bibfnamefont {X.}~\bibnamefont
  {Zhang}},\ }\bibfield  {title} {\bibinfo {title} {{Towards a systematic study
  of non-thermal leptogenesis from inflaton decays}},\ }\href
  {https://doi.org/10.1007/JHEP05(2024)147} {\bibfield  {journal} {\bibinfo
  {journal} {JHEP}\ }\textbf {\bibinfo {volume} {05}},\ \bibinfo {pages}
  {147}},\ \Eprint {https://arxiv.org/abs/2311.05824} {arXiv:2311.05824
  [hep-ph]} \BibitemShut {NoStop}%
\bibitem [{\citenamefont {Bezrukov}\ and\ \citenamefont
  {Shaposhnikov}(2008)}]{Bezrukov:2007ep}%
  \BibitemOpen
  \bibfield  {author} {\bibinfo {author} {\bibfnamefont {F.~L.}\ \bibnamefont
  {Bezrukov}}\ and\ \bibinfo {author} {\bibfnamefont {M.}~\bibnamefont
  {Shaposhnikov}},\ }\bibfield  {title} {\bibinfo {title} {{The Standard Model
  Higgs boson as the inflaton}},\ }\href
  {https://doi.org/10.1016/j.physletb.2007.11.072} {\bibfield  {journal}
  {\bibinfo  {journal} {Phys. Lett. B}\ }\textbf {\bibinfo {volume} {659}},\
  \bibinfo {pages} {703} (\bibinfo {year} {2008})},\ \Eprint
  {https://arxiv.org/abs/0710.3755} {arXiv:0710.3755 [hep-th]} \BibitemShut
  {NoStop}%
\bibitem [{\citenamefont {Lozanov}\ and\ \citenamefont
  {Amin}(2017)}]{Lozanov:2016hid}%
  \BibitemOpen
  \bibfield  {author} {\bibinfo {author} {\bibfnamefont {K.~D.}\ \bibnamefont
  {Lozanov}}\ and\ \bibinfo {author} {\bibfnamefont {M.~A.}\ \bibnamefont
  {Amin}},\ }\bibfield  {title} {\bibinfo {title} {{Equation of State and
  Duration to Radiation Domination after Inflation}},\ }\href
  {https://doi.org/10.1103/PhysRevLett.119.061301} {\bibfield  {journal}
  {\bibinfo  {journal} {Phys. Rev. Lett.}\ }\textbf {\bibinfo {volume} {119}},\
  \bibinfo {pages} {061301} (\bibinfo {year} {2017})},\ \Eprint
  {https://arxiv.org/abs/1608.01213} {arXiv:1608.01213 [astro-ph.CO]}
  \BibitemShut {NoStop}%
\bibitem [{\citenamefont {Gondolo}\ and\ \citenamefont
  {Gelmini}(1991)}]{Gondolo:1990dk}%
  \BibitemOpen
  \bibfield  {author} {\bibinfo {author} {\bibfnamefont {P.}~\bibnamefont
  {Gondolo}}\ and\ \bibinfo {author} {\bibfnamefont {G.}~\bibnamefont
  {Gelmini}},\ }\bibfield  {title} {\bibinfo {title} {{Cosmic abundances of
  stable particles: Improved analysis}},\ }\href
  {https://doi.org/10.1016/0550-3213(91)90438-4} {\bibfield  {journal}
  {\bibinfo  {journal} {Nucl. Phys. B}\ }\textbf {\bibinfo {volume} {360}},\
  \bibinfo {pages} {145} (\bibinfo {year} {1991})}\BibitemShut {NoStop}%
\end{thebibliography}%

\end{document}